\documentclass[preprintnumbers, superscriptaddress, showpacs, nofootinbib, amsfonts, amsmath, amssymb, aps, prd, notitlepage, floatfix]{revtex4-2}

\usepackage{graphicx}
\usepackage{curves}
\usepackage{hyperref}
\usepackage{amsmath}
\usepackage{graphicx}
\usepackage{bbold}
\usepackage{amssymb}
\usepackage{mathtools}
\usepackage{physics}
\usepackage{slashed}
\usepackage{xcolor}
\usepackage{MnSymbol}

\usepackage{subcaption}

\usepackage{amssymb}
\usepackage{lipsum}

\newcommand{\thalf}{\tfrac{1}{2}}

\begin{document}

%
\title{Probing Composite Structure and Spin-Orbit Coupling with GPDs in ${}^4$He}
%

\author{Antonio Garcia Vallejo}
\email[Email: ]{antoniov@nmsu.edu}
\affiliation{Department of Physics, New Mexico State University, Las Cruces, New Mexico 88003, USA}
\author{Matthew~D.~Sievert}
\email[Email: ]{msievert@nmsu.edu}
\affiliation{Department of Physics, New Mexico State University, Las Cruces, New Mexico 88003, USA}

\begin{abstract}

In this work, we extend the Impulse Approximation for the generalized parton distributions (GPDs) of a spin-0 composite hadron with spin-$\tfrac{1}{2}$ constituents to manifestly incorporate the symmetries of the target wave function.
The method utilizes a light-front Wigner function representation instead of a spectral density and a basis of Pauli matrices for the spin. It exploits the boost invariance of the light front and rotational invariance in the rest frame to parameterize the Wigner density in terms of only 3 structures.
In addition to the isotropic term and a term describing $\vec{L} \cdot \vec{S}$ coupling previously observed for transverse momentum dependent parton distributions (TMDs), we also identify a novel coupling $\Delta\vec{L}\cdot\vec{S}$ to the angular momentum \textit{transfer} which is unique to the case of GPDs.
We then apply the framework to a composite ${}^4$He target with simple phenomenological models to identify qualitative experimental signatures of composite-structure effects in light nuclei.  
The framework we have constructed here can be readily extended to the case of generalized TMDs (GTMDs), while the phenomenological framework is applicable both to analyses of light nucleus data and as training input for AI-assisted applications.

\end{abstract}


\date{\today}
\maketitle


\section{Introduction}
\label{introduction}

A central goal of hadronic physics is to understand the composition of hadrons and nuclei in terms of their constituents, and how the collective properties of the composite parent arise, such as its mass, charge, and spin \cite{JaffeManohar1990,Ji1997PRL,Ji1997PRD,Radyushkin1997,Mueller1994,JiLebed2001}.  In recent years, a major focus has been the study and extraction of distributions which characterize the three-dimensional (3D) spatial structure of the parton constituents of hadrons, known as hadron tomography \cite{Malace2014IJMPE,Seely2009PRL,Weinstein2011PRL,Hen2017RMP,Alrashed2022PRL,Kovchegov2019JHEP,Hatta2017PRD}. Of particular interest for this work are the Generalized Parton Distributions (GPDs) which may be extracted from exclusive processes like Deeply Virtual Compton Scattering (DVCS). The GPDs characterize the dependence of the hadronic matrix element on the momentum transfer $\Delta^\mu$ \cite{CollinsFrankfurtStrikman1997,CollinsFreund1999,BelitskyMuellerKirchner2002}, which is in turn related to the spatial distribution of probed partons through a Fourier transform.  The result, in principle, is hadron tomography: a visual representation (image) of the partonic densities inside the hadron of interest \cite{Burkardt2000,Burkardt2003,PireSofferTeryaev1999,Pobylitsa2002a,Pobylitsa2002b}.
Extracting the GPDs of the nucleon are of particular interest because GPD sum rules are one of the only known ways to set constraints on the orbital angular momentum of the nucleon \cite{Aidala:2013RMP,Adams:2024pxw}.  Understanding from first principles how the spin $\tfrac{1}{2}$ of the nucleon is decomposed in terms of its quark and gluon constituents is a decades-long puzzle in the field \cite{Ashman1988,JaffeManohar1990,Ji1997,Aidala2013,LeaderLorce2014,HattaZhao2018,Alexandrou2020}, and the contribution from orbital angular momentum remains the most elusive to pin down. 


Although GPDs are a promising tool for directly imaging the spatial structure of hadrons, in practice they are very difficult to extract from currently available exclusive scattering data \cite{Adams:2024pxw,Moutarde:2019CFF,Kumericki:2019Uncert,KumerickiMueller:2015,Shiells:2022JHEP,KriestenLiuti:2022PRD,Cuic:2020PRL}. One major direction of emphasis is the utilization of modern AI and machine learning techniques to assist in solving the GPD inverse problem; significant steps in this direction have already been taken by the EXCLAIM Collaboration \cite{Liuti:2024_EXCLAIM_Project,Liuti:2024_ML_Polarized,Almaeen:2024_VAIMCFF,Hossen:2024_CVAIM,Freese:2024_KernelGPD,dotson2025generalizedpartondistributionssymbolic}.  Current experimental data on GPDs comes primarily from the Thomas Jefferson National Accelerator Facility (JLab) and is limited in reach by the kinematics of the accelerator \cite{Stepanyan2001PRL, MunozCamacho2006PRL, Christiaens2023PRL, Hattawy2017PRL, Hattawy2019PRL, Dupre2021PRC}. However, the advent of the electron-ion collider (EIC) around 2035 will present a wealth of new data on exclusive scattering with the nucleon. That makes developments in GPD theory and phenomenology timely and crucial to develop and test the methods on JLab data, while preparing for future EIC data.

Notably, the EIC will perform high-precision imaging not only of the nucleon, but also of composite nuclei \cite{Accardi2016EPJA,EICYellow2022}.  Compared to the nucleon, the partonic structure of composite nuclei is little understood.  For one-dimensional parton distribution functions, nuclear effects can be parameterized through a parton distribution function (PDF) ratio (the nuclear modification factor) to study the one-dimensional modulation of the nucleon structure in a nuclear environment \cite{EPS09,DSSZ12,nCTEQ15,EPPS16,nNNPDF20,KlasenPaukkunen2024AR}. The EMC effect \cite{EMC1983,Arneodo1994,Geesaman1995,Accardi2016EPJA}, for instance, is a decades-long mystery reflecting the modification of the nucleon's partonic structure and associated short-range correlations \cite{Weinstein2011PRL,Sargsian2012,Schmookler2019Nature,Segarra2021PRR,ArringtonFominSchmidt2022}.  More differential observables in composite nuclei, like transverse momentum dependent parton distributions (TMDs) and GPDs, carry much more information about the internal hadronic structure. Whereas nuclear PDFs (one-dimensional) effectively view the nucleus as a static density of nucleons, observables like nuclear TMDs are sensitive even to the internal orbital angular momentum of the nucleons inside the nucleus \cite{Kovchegov:2015zha}. Of course, data on the three-dimensional partonic structure of nuclei is even more limited than for the nucleon, but current experiments like E12-17-012 using the ALERT detector at JLab will provide insight into the GPDs of light nuclei even before the arrival of the EIC \cite{Charles2015ALERT,ALERTRunGroup2016}.

With new data from JLab and the prospect of high-precision tomographic data on the nucleon and nuclei on the horizon from the EIC, there is a timely opportunity to advance the theory of GPDs for composite hadonic targets.  The most natural and literal interpretation of this would be to describe nuclear GPDs, such as light nuclei like ${}^4$He studied in experiment E12-17-012 with ALERT. However, composite models may also be appropriate for other hadronic systems, even without the clear distinction between nuclear and nucleon degrees of freedom.  For instance, one may investigate possible flavor asymmetries in the quark sea in a non-perturbative ``pion cloud'' model of the proton \cite{Shigetani1993,CaoSignal2003,HeWang2019,LuanLu2022}. In any case, GPDs viewed through the lens of the composite structure provide an excellent phenomenological laboratory to study the effects of orbital angular momentum and its coupling to spin -- as we demonstrate in this paper.

The main theoretical framework we employ here is the impulse approximation (IA), as implemented on the light front.  The impulse approximation was originally developed by Frankfurt and Strikman \cite{Frankfurt:1981mk,Frankfurt:1988} and has been applied to the study of nuclear GPDs by several groups \cite{Guzey:2003,Scopetta:2004,Scopetta:2009,Fucini:2020,Cosyn:2018} and to ${}^4\mathrm{He}$ in particular \cite{Liuti:2005}.  These approaches derive a master formula in terms of an off-forward product of nuclear wave functions, which is sometimes re-expressed as a spectral density.  In this work, we substantially extend this formalism by transforming to a Wigner density representation and making manifest the unbroken symmetries of the nuclear wave function.  In doing so, we obtain significant constraints on the form of the mixing between nuclear and nucleon matrix elements and parameterize the fundamental structures which can encode spin-orbit correlations in the composite nucleus.

The rest of this paper is organized as follows.  In Section~\ref{sec:composite-theory}, we present the theoretical derivation of the master formula for GPDs of a spin-0 composite hadron in terms of the GPDs of its spin-$\tfrac{1}{2}$ constituents.  In Eq.~\eqref{eq:GPD-momentum-decomp} we decompose the matrix elements of the composite hadron in terms of the matrix elements of its constituents, together with their Wigner density in the composite hadron.  We employ a Pauli matrix decomposition \eqref{eq:GPD-momentum-spin-decomp} for the spin degree of freedom and impose explicit rotational, parity, and time reversal invariance on the Wigner function, constraining its parameterization with the help of the dictionary \eqref{eq:z-boosted-quantities} between the target rest frame and the boosted frame.  The result is the master formula \eqref{eq:scalar-gpd-master} for the unpolarized quark GPD of the composite target in terms of the spin-dependent constituent GPDs and the parameterized Wigner distribution.  In the process, we identify a new type of $\Delta\vec{L} \cdot \vec{S}$ spin-orbit coupling \eqref{eq:wigner-pol-decomp} that contributes to composite-hadron GPDs which did not arise for the TMD case \cite{Kovchegov:2015zha}.

To illustrate the physical content of the master formula \eqref{eq:scalar-gpd-master}, in Sec.~\ref{sec:pheno} we consider a simple phenomenological model for composite ${}^4$He target, using the quark target model as a perturbative input for the constituent GPDs. While some aspects of the phenomenology are unique to the quark target model, we identify multiple generic signatures of composite-structure effects in the GPDs.  The resulting structure of the composite is a nontrivial combination of the spatial Fourier transform of the target Wigner density and longitudinal momentum smearing.  The presence of significant spin-orbit coupling in the composite target leads to systematic changes in the GPD, which we use to suggest possible experimental signatures of spin-orbit coupling.

We conclude in Sec.~\ref{sec:Concl} by summarizing the general versus model-specific features of our phenomenology and surveying natural next steps.  The methodology presented here can be readily extended to other theoretical cases, notably GTMDs, while the developed phenomenological framework can be directly applied to experimental data on light nuclei and also adapted as training data for the EXCLAIM Collaboration's AI-assisted phenomenology.  Additional calculations are presented in the Appendix \ref{ap:polarizations}, including the consistency check that the longitudinally-polarized quark GPD of a scalar target with composite structure vanishes explicitly, and a master integral similar to (\ref{eq:scalar-gpd-master}) for the distribution of transversely polarized quarks in the scalar composite system.


\section{Theory of composite structure for Quark GPDs of a scalar target}
\label{sec:composite-theory}

\subsection{Momentum decomposition}

The leading-twist quark GPDs in an unpolarized target are parametrized from the matrix element
\begin{align}
    \label{eq:quark-gpd-scalar}
	F_{q/\phi}^{[\Gamma]}
	&=
	\frac{1}{4\pi}\int dr^{-} e^{ixP^{+}r^{-}}
	(\Gamma)_{JI}\times 
	{}_{\text{out}}
	\bra{p'}\bar{\psi}_{J}(-r^{-}/2) \: \psi_{I}(r^{-}/2)\ket{p}{}_{\text{in}}
    \: ,
\end{align}
where $p$ and $p'$ are the initial (in) and final (out) momentum of the target hadron, $I, J$ are Dirac indices, and $\Gamma$ is a projection operator for the quark polarizations.  The expression \eqref{eq:quark-gpd-scalar} is valid in the light-front gauge $A^+ = 0$, whereas a general gauge contains a Wilson line connecting the quark fields.  The longitudinal momentum fraction carried by the struck quark is identified by $x$, the average momentum of the target is given by $P\equiv(p+p')/2$, and the momentum transfer is $\Delta = p' - p$. Note that we are free too choose a frame such that $\vec{P}_\bot = 0$ without loss of generality.  In the matrix element \eqref{eq:quark-gpd-scalar}, the quark field operators are separated by a light-front displacement $r^{-}$ \cite{Diehl2003,Mei_ner_2007} but evaluated at equal light-front time $r^+ = 0$ (as well as at $\vec{r}_\bot = 0$).  Here we utilize light-front coordinates with the convention
\begin{align}
    \label{eq:lf-convention}
    v^{\pm} 
    \equiv
    \frac{1}{\sqrt{2}} ( v^{0} \pm v^{3} )
    \: .
\end{align}

Now suppose that the target hadron whose GPDs are described by \eqref{eq:quark-gpd-scalar} is composite, with an intermediate constituent degree of freedom.  While the matrix element \eqref{eq:quark-gpd-scalar} is nonperturbative, it can be decomposed in terms of a complete set of Fock states corresponding to the free, on-shell constituents:
%
\begin{align}
    \label{eq:complete-set}
    \hat{\mathbb{1}}
	&= 
	\sumint\limits_{X} \ket{X} \bra{X}
    = \sum_n \frac{1}{S_n} \int\frac{d^2 p_{1\bot} \, dp_1^+}{(2\pi)^3 2 p_1^+} \cdots \frac{d^2 p_{n \bot} \, dp_n^+}{(2\pi)^3 2 p_n^+} \: \ket{p_1 \cdots p_n} \bra{p_1 \cdots p_n}
    \: ,
\end{align}
where $n$ denotes the number of constituent particles and $S_n$ is a symmetry factor for identical particles.  The sum also runs over the types and polarizations of constituents (not explicitly shown for brevity).

First we decompose the in and out states using \eqref{eq:complete-set}, which reduces the matrix element \eqref{eq:quark-gpd-scalar} to simpler amplitudes that can be calculated in terms of the constituent degrees of freedom:
\begin{align}
    \label{eq:matrix-decomp}
	{}_{\text{out}} 
	\bra{p'}\bar{\psi}_{J}\big(\frac{-r^{-}}{2}\big)  \psi_{I}\big(\frac{r^{-}}{2}\big)\ket{p}{}_{\text{in}}
	&=
	\sumint\limits_{\left\{ X, Y \right\}}
	{}_{\text{out}}
	\bra{p'}\ket{X}{}_{\text{out}} \, {}_{\text{out}}\bra{X}
	\bar{\psi}_{J}(-r^{-}/2) \psi_{I}(r^{-}/2)
	\ket{Y}{}_{\text{in}} \, {}_{\text{in}}\bra{Y}\ket{p}{}_{\text{in}}
    \: .
\end{align}
Due to Lorentz invariance, the projection of the hadronic state $\ket{p}_\mathrm{in}$ onto the constituent Fock state $\ket{Y}_\mathrm{in}$ must be proportional to an overall momentum-conserving delta function, 
\begin{align}   
    \label{e:WFdeltafn1}
    {}_{\text{in}} \bra{ Y }\ket{p}{}_{\text{in}} 
    &\equiv
    (2\pi)^3 \, 2 p^+ \: \delta^2 \bigg( \sum_i \vec{p}_{Y i \bot} - \vec{p}_{\bot} \bigg) \,
    \delta\bigg( \sum_i p_{Y i}^+ - p^+\bigg) \:
    \Psi(
    \{ p_{Y i}^+  \} \: ; \:
    \{\vec{p}_{Y i\bot} \} ) \: ,
\end{align}
with the amplitude $\Psi$ that remains giving the multiparticle light-front wavefunction (LFWF) of the constituents within the composite target.  Moreover, unlike the instant-form wave function, Lorentz invariance of the amplitude \eqref{e:WFdeltafn1} implies that the LFWF is also \textit{boost invariant} and Galilean invariant on the light front.  As a result, the LFWF 
\begin{align}
    \Psi\left( \{ p_{Y i}^+  \} \: ; \: \{\vec{p}_{Y i\bot} \} \right)
    =
    \Psi\bigg( \left\{ \frac{p_{Y i}^+}{p^+}  \right\} \: ; \: \{\vec{p}_{Y i\bot} \} \bigg) 
\end{align}
depends only on the longitudinal momentum fractions $x_i = p_i^+ / p^+$ and intrinsic transverse momenta $\vec{p}_{i\bot}$ of the constituents (as well as their polarizations).
Detailed reviews can be found in Refs.~\cite{Kovchegov:2012mbw, Lepage:1980fj, Brodsky:1997de, Weinberg:1995mt}. 

Additionally, of the various constituent particles entering the Fock state matrix element $\bra{X} \bar{\psi}(-r^{-}/2) \psi(r^{-}/2) \ket{Y}$, only a subset of the particles will be connected to the hard-scattering vertex represented by the quark operators $\bar\psi \: \psi$.  The remaining spectator particles are, by definition, disconnected from the amplitude and factor out of it:
\begin{align}   
    \label{e:spectators1}
    \bra{X} \bar{\psi}_{J}(-r^{-}/2) \psi_{I}(r^{-}/2) \ket{Y} &=
    \bra{ \left\{P_{n}'\right\} } \bar{\psi}_{J}(-r^{-}/2) \psi_{I}
     (r^{-}/2) \ket{ \left\{P_{n}\right\} } \: \times \:
    \braket{X'}{Y'}
    \: ,
\end{align}
where we have introduced the notation in which we denote the subset of partons from the intermediate states $\left\{ X, Y \right\}$ which are spectators as $\left\{ X' , Y' \right\}$, while the active constituents (connected to the hard scattering)  are labeled $\left\{ P_{n} , P_{n}' \right\}$.  

Using Eqs.~\eqref{e:WFdeltafn1} and \eqref{e:spectators1} in \eqref{eq:matrix-decomp} gives
\begin{align}
    \label{e:matrix-decomp-conn-spec}
	{}_{\text{out}} 
	\bra{p'}\bar{\psi}_{J}\big(\frac{-r^{-}}{2}\big)  \psi_{I}\big(\frac{r^{-}}{2}\big)\ket{p}{}_{\text{in}}
	&=
	\sumint\limits_{\left\{ X, Y \right\}}
    (2\pi)^3 \, 2 p^{\prime \: +} \: \delta^2 \bigg( \sum_i \vec{p}_{X i \bot} - \vec{p}_{\bot}^{\: \prime} \bigg) \,
    \delta\bigg( \sum_i p_{X i}^+ - p^{\prime \: +}\bigg) \:
    \Psi^*(X; p')
    \notag 
    \\ 
    &\quad\times
    (2\pi)^3 \, 2 p^{+} \: \delta^2 \bigg( \sum_j \vec{p}_{Y j \bot} - \vec{p}_{\bot}\bigg) \,
    \delta\bigg( \sum_j p_{Y j}^+ - p^{+}\bigg) \:
    \Psi(Y; p)
    \notag 
    \\ 
    &\qquad\times
	\bra{ \left\{P_{n}'\right\} } \bar{\psi}_{J}(-r^{-}/2) \psi_{I}
     (r^{-}/2) \ket{ \left\{P_{n}\right\} } \: 
    \braket{X'}{Y'} \: .
\end{align}
%


\noindent
The scattering matrix for the passive spectators among themselves, which is disconnected from the quark field operators, again is proportional to an overall momentum-conserving delta function for the spectators:
\begin{align}
    \label{eq:spectator-s-matrix}
    \braket{X'}{Y'}
    &=
    {}_{\text{out}}\bra{ \left\{ ( p_{X'} , s_{X'} )  \right\}} \ket{ \left\{ ( p_{Y'} , s_{Y'} )  \right\} }{}_{\text{in}}
    \notag
    \\ &=
    (2\pi)^3 2 P_{X', tot}^+ \, \delta\left( \sum_{i} p_{X' \, i}^{+} - \sum_{i}p_{Y' \, i}^{+} \right) 
    \delta^{2} \left( \sum_{i}\vec{p}_{X' \, i\perp } - \sum_{i}\vec{p}_{Y' \, i\perp } \right) 
    \mathcal{S}_{\left\{s_{Y'}\right\}}^{\left\{s_{X'}\right\}}\left( \left\{p_{X'}\right\} , \left\{p_{Y'}\right\} \right) 
    \: ,
\end{align}
%
%
where $P_{X', tot}^+$ is the total longitudinal momentum from the spectator states.
%

\begin{figure}
    \centering
    \includegraphics[width=0.6\linewidth]{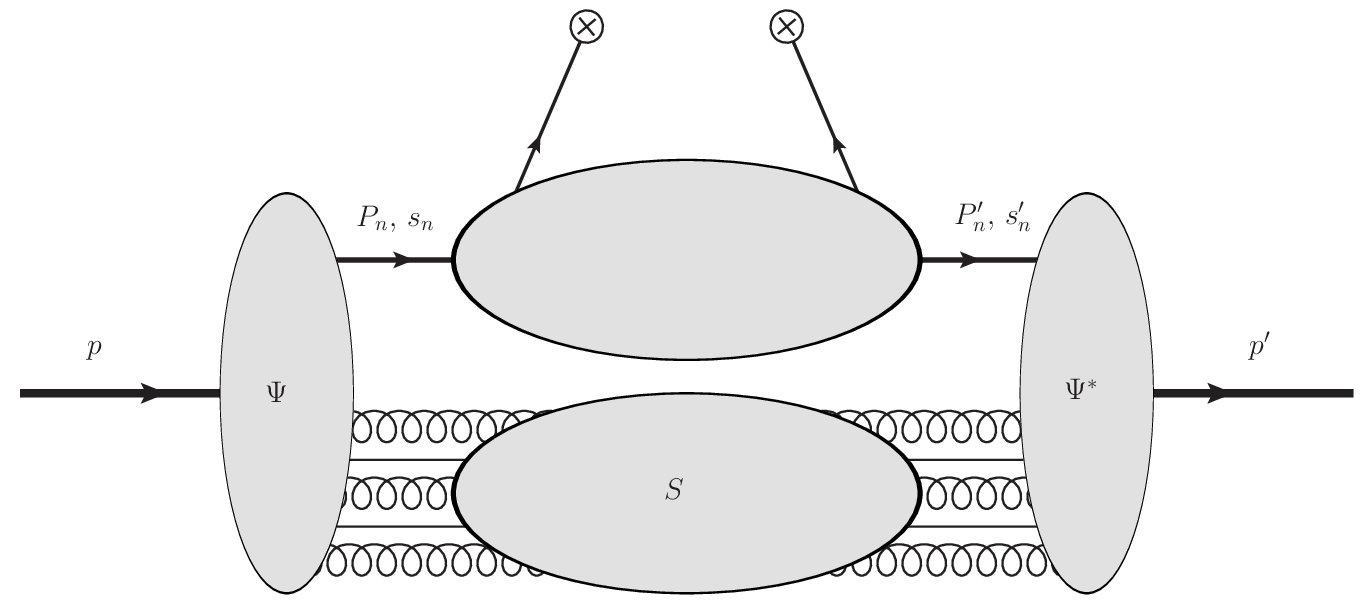}
    \caption{Illustration of the decomposition \eqref{eq:matrix-decomp}.  The overall quark GPD of the scalar composite target is decomposed into four distinct components: the wave functions $\Psi^*, \Psi$ of the constituents, the scattering amplitude $\mathcal{S}$ of the spectators, and the quark matrix element of the active (connected) constituents.  In the mean-field approximation, only one active constituent is considered at a time.}
    \label{fig:scalar-gpd-decomp}
\end{figure}

Now we combine the previous expressions to simplify the quark matrix element; the structure of the expansion is illustrated in Fig.~(\ref{fig:scalar-gpd-decomp}).  
Enforcing momentum conservation explicitly with the aid of the delta functions in (\ref{e:matrix-decomp-conn-spec}) and (\ref{eq:spectator-s-matrix}) allows us to decompose the quark matrix element as 
\begin{align}
    \label{eq:matrix-decomp-mfa}
    {}_{\text{out}} 
	\bra{p'}\bar{\psi}_{J}\big(\frac{-r^{-}}{2}\big)  \psi_{I}\big(\frac{r^{-}}{2}\big)\ket{p}{}_{\text{in}}
     &=
     \sumint\limits_{ \left\{ P_{n} , P_{n}' \right\} }
     \sumint\limits_{ \left\{  X' , Y' \right\} }
     \bra{ \left\{P_{n}'\right\} } \bar{\psi}_{J}(-r^{-}/2) \psi_{I}
     (r^{-}/2) \ket{ \left\{P_{n}\right\} } \:
     \mathcal{S}(X', Y') \:
     \Psi^{*}(X', P_n^\prime ; p') \: 
     \Psi(Y', P_n ; p)
     \notag \\ &\quad\times
     (2\pi)^3 2 P_{X', tot}^+ \,  
    \delta^{2} \left( \sum_{i}\vec{p}_{X' \, i\perp } - \sum_{i}\vec{p}_{Y' \, i\perp } \right) 
    \delta\left( \sum_{i} p_{X' \, i}^{+} - \sum_{i}p_{Y' \, i}^{+} \right)
     \notag \\ &\quad\times 
     (2\pi)^3 \, 2 p^{\prime \: +} \: \delta^2 \bigg( \sum_i \vec{p}_{X ' i \bot} + \sum_{i} \vec{P}_{i}' - \vec{p}_{\bot}^{\: \prime} \bigg) \,
    \delta\bigg( \sum_i p_{X ' i}^+  + \sum_{i} P_{i}^{\prime \: +} - p^{\prime \: +}\bigg) 
    \notag 
    \\ 
    &\quad\times
    (2\pi)^3 \, 2 p^{+} \: \delta^2 \bigg( \sum_j \vec{p}_{Y ' j \bot} + \sum_{j} \vec{P}_{j} - \vec{p}_{\bot}\bigg) \,
    \delta\bigg( \sum_j p_{Y ' j}^+ + \sum_{j}^{n} P_{j}^{+} - p^{+}\bigg)
    \notag \\ \notag \\ &=
    \sumint\limits_{ \left\{ P_{n} , P_{n}' \right\} }
    \bra{ \left\{P_{n}'\right\} } \bar{\psi}_{J}(-r^{-}/2) \psi_{I}
     (r^{-}/2) \ket{ \left\{P_{n}\right\} } \:
    \notag \\ &\hspace{1cm}\times
    (2\pi)^3 \, 2 p^{\prime \: +} \: 
    \delta^2 \bigg( 
    \sum_{i} \vec{P}_{i}' - \sum_{j} \vec{P}_{j}  - \vec{\Delta}_{\bot}
    \bigg) \,
    \delta\bigg( 
    \sum_{i} P_{i}^{\prime \: +} 
    - \sum_{j} P_{j}^{+} 
    - \Delta^{+}  \bigg) 
    \notag \\ &\times
    \Bigg[
    \sumint\limits_{ \left\{  X' , Y' \right\} }
    \Psi^{*}(X', P_n^\prime ; p') \: 
     \Psi(Y', P_n ; p) \: 
    \mathcal{S}(X', Y') 
    \notag \\ &\quad\times 
    (2\pi)^3 2 P_{X', tot}^+ \, 
    \delta^{2} \left( \sum_{i}\vec{p}_{X' \, i\perp } - \sum_{i}\vec{p}_{Y' \, i\perp } \right)  
    \delta\left( \sum_{i} p_{X' \, i}^{+} - \sum_{i}p_{Y' \, i}^{+} \right) 
    \notag \\ &\quad\times 
    (2\pi)^3 \, 2 p^{+} \: 
    \delta^2 \bigg( \sum_j \vec{p}_{Y ' j \bot} + \sum_{j} \vec{P}_{j} - \vec{p}_{\bot}\bigg) \,
    \delta\bigg( \sum_j p_{Y ' j}^+ + \sum_{j} P_{j}^{+} - p^{+}\bigg)
    \Bigg]
    \notag \\ \notag \\ &=
    (2\pi)^3
    \sumint\limits_{ \left\{ P_{n} , P_{n}' \right\} }
    \frac{p^{+}}{P^+}
    \left(\frac{p^{+} - \sum_{j} P_{j}^{+}}{P^+}  \right) \:\:
    \tilde\Psi^*(\{ P_n^\prime \}; p') \: \tilde\Psi(\{ P_n \} ; p)
    \notag \\ &\hspace{1cm}\times
    \bra{ \left\{P_{n}'\right\} } \bar{\psi}_{J}(-r^{-}/2) \psi_{I}
    (r^{-}/2) \ket{ \left\{P_{n}\right\} } 
    \notag \\ &\hspace{1cm}\times
    (2\pi)^3 \, 2 p^{\prime \: +} \: 
    \delta^2 \bigg( 
    \sum_{i} \vec{P}_{i}' - \sum_{j} \vec{P}_{j}  - \vec{\Delta}_{\bot}
    \bigg) \,
    \delta\bigg( 
    \sum_{i} P_{i}^{\prime \: +} 
    - \sum_{j} P_{j}^{+} 
    - \Delta^{+}  \bigg) \: ,
\end{align}
where we have defined the inclusive wave functions 
\begin{align}
    \tilde\Psi^*(\{ P_n \}; p') \: 
    \tilde\Psi(\{ P_n^\prime \}; p) 
    &\equiv
    \frac{1}{2} (2\pi)^3
    \sumint\limits_{ \left\{  X' , Y' \right\} }
    \Psi^{*}(X', P_n^\prime ; p') \: 
    \Psi(Y', P_n ; p) \: 
    \mathcal{S}(X', Y') 
    \notag \\ &\quad\times 
    2 P^+  \, 
    \delta^{2} \left( \sum_{i}\vec{p}_{X' \, i\perp } - \sum_{i}\vec{p}_{Y' \, i\perp } \right)  
    \delta\left( \sum_{i} p_{X' \, i}^{+} - \sum_{i}p_{Y' \, i}^{+} \right) 
    \notag \\ &\quad\times 
    2 P^+ \: 
    \delta^2 \bigg( \sum_j \vec{p}_{Y ' j \bot} + \sum_{j} \vec{P}_{j} - \vec{p}_{\bot}\bigg) \,
    \delta\bigg( \sum_j p_{Y ' j}^+ + \sum_{j} P_{j}^{+} - p^{+}\bigg)
\end{align}
by integrating out the spectators.  The constants have been chosen for later convenience.

In general Eq.~\eqref{eq:matrix-decomp-mfa} contains a sum over all numbers of active (connected) constituents.  The sum is an expansion in the number of correlated constituents within the composite hadron, with the simplest term being the case of one connected constituent.  Keeping only the one-particle term constitutes a mean field approximation which is referred to as the impulse approximation (IA) \cite{Guzey:2003,Scopetta:2004,Scopetta:2009,Fucini:2020,Cosyn:2018,Liuti:2005}. In principle, the mean-field IA is just the first term of an infinite series expansion in the correlated constituents, which we may formally write as
\begin{align}   \label{e:MFexpand1}
    {}_{\text{out}} 
	\bra{p'}\bar{\psi}_{J}\big(\frac{-r^{-}}{2}\big)  \psi_{I}\big(\frac{r^{-}}{2}\big)\ket{p}{}_{\text{in}}
    &=
    \frac{1}{2} \sum_{s s'}
    \int\frac{d^2 P_{1\bot} \, dP_1^+}{2 P_1^+} \:
    \frac{d^2 P_{1\bot}^\prime \: dP_1^{\prime \: +}}{(2\pi)^3 \, 2 P_1^{\prime \: +}} \:\:
    \frac{p^{+}}{P^+}
    \left(\frac{p^{+} - P_1^+}{P^+}  \right) \:\:
    \tilde\Psi_{s'}^*(P_1^\prime; p') \: \tilde\Psi_s( P_1 ; p)
    \notag \\ &\hspace{1cm}\times
    \bra{ P_1' s' } \bar{\psi}_{J}(-r^{-}/2) \psi_{I}
    (r^{-}/2) \ket{ P_1 s } 
    \notag \\ &\hspace{1cm}\times
    (2\pi)^3 \, 2 p^{\prime \: +} \: 
    \delta^2 \bigg( \vec{P}_1^\prime - \vec{P}_1  - \vec{\Delta}_{\bot} \bigg) \,
    \delta\bigg( P_1^{\prime \: +} - P_1^{+} - \Delta^{+}  \bigg) 
    \notag \\ &\hspace{1cm}+
	\text{correlations} 
    \notag \\ \notag \\ &=
    \frac{1}{2} \sum_{s s'}\int d^{2}P_{1\perp} dP_{1}^{+}
    \Bigg(
    \frac{1}{2 P_{1}^{+}}
    \frac{p^{\prime \: +}}{P_{1}^{+} + \Delta^{+} } 
    \frac{p^{+}}{ P^{+} }
    \frac{ p^{+} - P_{1}^{+} }{ P^{+} }
    \Bigg)
    \tilde{\Psi}_{s'}^{*}( P_{1} + \Delta  ; p' ) \: \tilde{\Psi}_s ( P_{1} ; p )
    \notag
    \\
    &\qquad\times
    \bra{ (P_1 + \Delta) s' }  \bar{\psi}_{J}(-r^{-}/2) \psi_{I}(r^{-}/2) 
    \ket{  P_{1}  s  }
    \: + \: 
    \text{correlations}  
\end{align}
Here $s, s'$ denote the spin quantum numbers of the constituents.  In the IA, a momentum conservation constraint immediately follows directly from this single-constituent approximation: 
\begin{align}
    P_{1}' - P_{1}  = \Delta = p' - p   \: .
\end{align}
A mean field approximation can be a good description of a variety of systems, such as heavy nuclei and perturbative QCD in the large-$N_c$ limit \cite{HOOFT1974461}.  Depending on the system and the observable, multiparticle correlations may be more or less important.

Up through Eq.~\eqref{e:MFexpand1}, the derivation of the IA convolution formula is essentially the same as what has been shown previously \cite{Guzey:2003,Scopetta:2004,Scopetta:2009,Fucini:2020,Cosyn:2018,Liuti:2005}.  To go beyond this representation, we need to transform the way the nuclear wave functions are represented.
Eq.~\eqref{e:MFexpand1} contains, as a fundamental ingredient, the product of two light-front wave functions with different kinematic arguments.  Such an off-forward product can be powerfully and intuitively represented through a \textit{Wigner transform}.  If we define the Fourier transform of the momentum-space wave function as
\begin{align}   \label{e:FTdefn}
\begin{aligned}
    \frac{\tilde\Psi_s(k^+, \vec{k}_\bot)}{\sqrt{2 k^+}} &\equiv
    \int d^2 b_\bot db^- \: e^{+ i k^+ b^- - i \vec{k}_\bot \cdot \vec{b}_\bot} \: \tilde\Psi_s(b^- , \vec{b}_\bot)   \:  ,
    \\
    \tilde\Psi_s(b^- , \vec{b}_\bot) &= \int\frac{d^2 k_\bot \, dk^+}{(2\pi)^3} \:
    e^{-i k^+ b^- + i \vec{k}_\bot \cdot \vec{b}_\bot} \:
    \frac{\tilde\Psi_s(k^+, \vec{k}_\bot)}{\sqrt{2 k^+}}  \: ,
\end{aligned}
\end{align}
so as to absorb the relativistic phase space normalization, then the Wigner function, defined by
\begin{align}   \label{e:WignerDefn}
    W_{s' s} ( 
    p^+ , \vec{p}_\bot \: ; \: 
    b^- , \vec{b}_\bot 
    )
	&\equiv 
	\int\frac{d^2 q_\bot \, dq^+}{(2\pi)^{3}} 
    e^{-i q^+ b^- + i \vec{q}_\bot \cdot \vec{b}_\bot} \:
	\frac{\tilde\Psi_{s'}^{*}( p - \frac{1}{2}q )}{ \sqrt{ 2 (p - \frac{1}{2}q)^+ } } 
	\frac{\tilde\Psi_s( p + \frac{1}{2}q )}{ \sqrt{ 2 (p + \frac{1}{2}q)^+ } } 
    \notag \\ &=
    \int d^2 r_\bot \, dr^-
    e^{+i p^+ r^- - i \vec{p}_\bot \cdot \vec{r}_\bot} \:
	\tilde\Psi_{s'}^{*}( b - \frac{1}{2}r ) \:
	\tilde\Psi_s( b + \frac{1}{2}r )
\end{align}
provides the fully quantum analog of the classical phase-space distribution in position and momentum.  Although the Uncertainty Principle forbids simultaneously localizing a wave function in both position and momentum space, the Wigner function \eqref{e:WignerDefn} circumvents that limitation at the level of the wave function \textit{squared}, by localizing only the \textit{average} position and momentum of the two wave functions.  This ability to interpret the Wigner function semi-classically makes it especially suitable for modeling, or for direct construction from first-principles wave functions.  With the definition \eqref{e:WignerDefn}, the Wigner function is normalized to unity (assuming the wave functions themselves are normalized to unity):
\begin{align}   \label{e:WignerNorm}
    \sum_{s s'} \, \delta_{s s'} \: 
    \int\frac{d^2 p_\bot dp^+}{(2\pi)^3} \, d^2 b_\bot db^- \:\:
    W_{s' s} (p,b) 
    = \sum_{s} \int\frac{d^2 p_\bot dp^+}{(2\pi)^3 2 p^+} 
    \left| \tilde\Psi_s(p) \right|^2
    = \sum_{s} \int d^2 b_\bot db^- \: 
    \left| \tilde\Psi_s(b) \right|^2
    = 1  \: .
\end{align}

Inverting the definition \eqref{e:WignerDefn} allows us to express any off-forward product of wave functions in terms of the Wigner function:
\begin{align}   \label{e:WignerTransform}
    \frac{\tilde\Psi_{s'}^{*}(k')}{ \sqrt{ 2 k^{\prime \, +} } } 
	\frac{\tilde\Psi_s(k)}{ \sqrt{ 2 k^+ } }
    =
    \int d^2 b_\bot \, db^- \: e^{+i (k - k')^+ b^- - i (\vec{k} - \vec{k}^{\: \prime})_\bot \cdot \vec{b}_\bot} \:\:
    W_{s' s} \Big( \thalf(k^+ + k^{\prime \:+}) , 
    \thalf(\vec{k}_\bot + \vec{k}^\prime_\bot) \: ; \: b^- , \vec{b}_\bot  \Big)    \: .
\end{align}
Employing the Wigner transform \eqref{e:WignerTransform} in \eqref{e:MFexpand1} allows us to further rewrite the quark matrix element in a composite hadron as
\begin{align}   
    {}_{\text{out}} 
	\bra{p'}\bar{\psi}_{J}\big(\frac{-r^{-}}{2}\big)  \psi_{I}\big(\frac{r^{-}}{2}\big)\ket{p}{}_{\text{in}}
    &=
    \frac{1}{2} \sum_{s s'} 
    \int d^{2}P_{1\perp} \, dP_{1}^{+}
    \int d^2 b_\bot \, db^- \:
    e^{-i \Delta^+ b^- + i \vec{\Delta}_\bot \cdot \vec{b}_\bot} \:\:
    \Bigg(
    \frac{1}{\sqrt{P_{1}^{+}}}
    \frac{p^{\prime \: +}}{\sqrt{P_{1}^{+} + \Delta^{+}} } 
    \frac{p^{+}}{ P^{+} }
    \frac{ p^{+} - P_{1}^{+} }{ P^{+} }
    \Bigg)
    \notag
    \\
    &\qquad\times
    W_{s' s} (P_1 + \thalf \Delta, b; P, \Delta) \:\:
    \bra{ (P_1 + \Delta) s' }  \bar{\psi}_{J}(-r^{-}/2) \psi_{I}(r^{-}/2) 
    \ket{  P_{1}  s  }
    \: + \: 
    \text{correlations}     \: .
\end{align}
Note that, because in this off-forward amplitude the two LFWF have composite parents with \textit{different} momenta ($p$ vs $p'$), the Wigner function acquires an additional dependence beyond that shown in Eq.~\eqref{e:WignerDefn} on the parent momenta $p, p'$ (or equivalently, $P, \Delta$).  Lastly, we define the average momentum $P_{N} := (P_{1} + P_{1}')/2$ of the active constituent and change variables from $P_{1} = P_{N} - \frac{\Delta}{2}$ to $P_N$.  We also introduce the skewness $\xi$, constituent momentum fraction $\alpha$, and partonic momentum fractions $x', \xi_N$ with respect to the active constituent:
\begin{align}
    \xi \equiv - \frac{\Delta^+}{2P^+}   
    \qquad , \qquad
    \alpha \equiv \frac{P_N^+}{P^+}
    \qquad , \qquad
    x' \equiv \frac{x P^+}{P_N^+} = \frac{x}{\alpha}
    \qquad , \qquad
    \xi_N \equiv - \frac{\Delta^+}{2P_N^+} = \frac{\xi}{\alpha}  \: .
\end{align}
Doing so gives
\begin{align}   
    {}_{\text{out}} 
	\bra{p'}\bar{\psi}_{J}\big(\frac{-r^{-}}{2}\big)  \psi_{I}\big(\frac{r^{-}}{2}\big)\ket{p}{}_{\text{in}}
    &=
    \frac{1}{2} \sum_{s s'} 
    \int d^{2}P_{N\perp} \, dP_{N}^{+} \:
    d^2 b_\bot \, db^- \:
    e^{-i \Delta^+ b^- + i \vec{\Delta}_\bot \cdot \vec{b}_\bot} \:\:
    \notag \\ &\times
    \Bigg(
    \frac{1}{\sqrt{ P_N^+ - \thalf \Delta^+ }}
    \frac{P^+}{\sqrt{ P_N^+ + \thalf \Delta^+ } } 
    \frac{P^+ + \thalf \Delta^+}{P^+}
    \frac{P^+ - \thalf \Delta^+}{ P^{+} }
    \frac{ P^+ - P_N^{+} }{ P^{+} }
    \Bigg)
    \notag \\ &\times
    W_{s' s} (P_N, b ; P, \Delta) \:\:
    \bra{ (P_N + \thalf \Delta) s' }  \bar{\psi}_{J}(-r^{-}/2) \psi_{I}(r^{-}/2) 
    \ket{  (P_N - \thalf \Delta)  s  }
    \: + \: 
    \text{correlations}
    \notag \\ \notag \\ &=
    \frac{1}{2} \sum_{s s'} 
    \int d\alpha \: d(P^+ b^-) \:
    d^2 P_{N\perp} \: d^2 b_\bot \:
    e^{-i \Delta^+ b^- + i \vec{\Delta}_\bot \cdot \vec{b}_\bot} \:\:
    \frac{1-\xi^2}{\sqrt{ 1  - \xi_N^2 }} \:
    \frac{1-\alpha}{\alpha} \:\:
    W_{s' s} (P_N, b) 
    \notag \\ &\times
    \bra{ (P_N + \thalf \Delta) s' }  \bar{\psi}_{J}(-r^{-}/2) \psi_{I}(r^{-}/2) 
    \ket{  (P_N - \thalf \Delta)  s  }
    \: + \: 
    \text{correlations}
    \: ,
\end{align}
so that the GPD matrix element \eqref{eq:quark-gpd-scalar} becomes
\begin{align}
    \label{eq:GPD-momentum-decomp}
	F_{q/\phi}^{[\Gamma]} (x, P, \Delta)
	&=
    \frac{1}{2} \sum_{s s'} 
    \int d\alpha \: d(P^+ b^-) \:
    d^2 P_{N\perp} \: d^2 b_\bot \:
    e^{-i [ \Delta^+ b^- -  \vec{\Delta}_\bot \cdot \vec{b}_\bot ]} \:\:
    \notag \\ &\times
    \frac{1-\xi^2}{\sqrt{ 1  - \xi_N^2 }} \:
    \frac{1-\alpha}{\alpha} \:\:
    W_{s' s} (P_N, b ; P, \Delta) \:\: 
    F_{q/N}^{[\Gamma]} \left(\tfrac{x}{\alpha}, P_N, \Delta ; s' s\right)
    \: + \: 
    \text{correlations}
    \: ,
\end{align}
The momentum decomposition found in Eq.~(\ref{eq:GPD-momentum-decomp}) is the first major theoretical result of this work. What it describes physically is the possibility to express the GPDs of an unpolarized (spin-0) hadronic target in terms of more fundamental GPDs constructed from the intermediate constituents, weighted by the corresponding Wigner distribution of those constituents.  The Wigner representation \eqref{eq:GPD-momentum-decomp} goes further than the wave function representation \eqref{e:MFexpand1} by re-expressing the (fully quantum) matrix element in a semi-classical language.  While the mean-field convolution structure is the same as in previous work, the Wigner representation \eqref{eq:GPD-momentum-decomp} provides a systematic framework for imposing symmetry constraints on the composite system.

The crucial achievement of the IA convolution \eqref{eq:GPD-momentum-decomp} is the separation of the ingredients which depend upon the properties of the hard scattering process (such as DVCS) which \textit{measures} the GPD from the \textit{intrinsic} properties of the wave function of the composite hadron itself.  We emphasize that \eqref{eq:GPD-momentum-decomp} holds for any spin constituents in a spin 0 (or unpolarized) composite target.  Mathematically, the quark correlator \eqref{eq:quark-gpd-scalar} from which the GPDs are defined has chosen the $z$-axis (light-front coordinates $P^+, r^-$) as a special direction in the problem.  The dependence on this auxiliary axis, sometimes denoted $n^\mu$, reflects the ``memory'' of the correlator about the hard-scattering process (like DVCS) which measured it.  That is, the hard scattering process is a function of both the target momentum $p^\mu$ and the virtual photon momentum $q_\gamma^\mu$.  The factorization of the process into a hard vertex and a GPD matrix element retains a dependence on the \textit{direction} of $n^\mu$, although not its kinematics.  This dependence on the auxiliary direction $n^{\mu}$ breaks symmetries such as 3D rotational invariance and time reversal invariance -- even if the underlying wave functions respect those symmetries.

The power of \eqref{eq:GPD-momentum-decomp} is that we can cleanly separate the parts of the process which depend on $n^\mu$ -- namely, the constituent GPDs $F^{[\Gamma]}_{q/N}$ -- from the parts which are $n$-independent -- namely the constituent Wigner function $W(p, b)$.  Since $W(p,b)$ is independent of $n^\mu$ (or the virtual photon kinematics $q_\gamma^\mu$), it is a function only of the kinematics of the composite hadron and its constituents.  Its symmetry properties are most transparent in the rest frame of the massive composite hadron, where it possesses full 3D rotational symmetry and unbroken discrete symmetries of parity and time reversal\footnote{That is not to say that the \textit{wave function} itself is invariant under all such symmetries; rather, various phases which accompany the transformation of the wave function cancel in the transformation of the Wigner density.}.

We note that there is an important subtlety regarding the impact parameter $b^\mu$.  The impact parameter $b$ defines a displacement from the center of light-front energy and does not transform as a proper four-vector  \cite{Lorc__2018,Lorc__2021}.  It is conceptually distinct from the displacement $\ell^\mu$ from the center of mass.  Strictly speaking, it is the $\ell^\mu$ dependence which exhibits rotational symmetry when boosted to the rest frame, not $b^\mu$.  Nevertheless, in the infinite-momentum frame $P^+ \rightarrow \infty$ in which the definition \eqref{eq:quark-gpd-scalar} holds, $b^\mu$ coincides with $\ell^\mu$, as discussed in Appendix~\ref{app:relativistic-centers}.  All formulas herein which relate to the three-dimensional properties of $\vec{b}$ are strictly valid for $\vec{\ell}$.  While we do not distinguish between $b$ and $\ell$ in the main text, we carefully explain the distinction in Appendix~\ref{app:relativistic-centers}.

Just as importantly, we intentionally constructed the Wigner distribution out of the Light-Front Wave Functions \eqref{e:WFdeltafn1} to ensure that $W(p,b)$ is \textit{manifestly boost invariant}.  A Wigner transform could be applied to any function, but since the LFWF are functions only of boost-invariant momentum fractions and not of $p^+$ directly, the light-front Wigner function \eqref{e:WignerDefn} is a boost-invariant function of boost-invariant quantities.  That is,
\begin{align}
    W_{s' s} ( P_N , b ; P, \Delta) = 
    W_{s' s} \left(\alpha, \vec{P}_{N \bot} 
    \: , \: P^+ b^- , \vec{b}_\bot 
    \: ; \: \xi, \vec{\Delta}_\bot 
    \right)  \: .
\end{align}
We will later exploit this crucial property by equating the Wigner function in the ``lab frame'' (with given $p, p'$ and $\vec{P}_\bot \equiv 0$) to the rest frame of the composite hadron, where $\vec{P} \equiv 0$.  Since these frames are connected by a pure boost along the $z$ axis, $W(p,b)$ is invariant:
\begin{align}   \label{e:W_invariance}
     W_{s's} (P_N, b; P, \Delta) \Big|_\mathrm{lab \: frame} 
     &=
     W_{s's} (\vec{P}_N, \vec{b}, \vec{\Delta}) \Big|_\mathrm{rest \: frame}
     %
     = 
     %
    W_{s' s} \left(\alpha, \vec{P}_{N \bot} 
    \: , \: P^+ b^- , \vec{b}_\bot 
    \: ; \: \xi, \vec{\Delta}_\bot 
    \right)
     \: .
\end{align}
The connection between $W(P_{N},b; P, \Delta)$ in the lab frame and the rest frame is invaluable, because in the rest frame, the Wigner function can be expanded through its tensorial dependence on the three-vectors $\vec{P}_{N}, \vec{b}, \vec{\Delta}$.  Although the underlying symmetry between the longitudinal and transverse directions is obscured when measured in the lab frame in terms of $p^+$ and $\vec{p}_\bot$, that symmetry is made manifest in the rest frame. The boost-invariant LF Wigner distributions \eqref{e:WignerDefn} make this connection possible.

In practice, one can take the quark matrix element $F^{[\Gamma]}_{q/N}$ for a known constituent and choose a model Wigner distribution $W(P_{N},b; P, \Delta)$ for the overall structure of the composite target, then use Eq.~\eqref{eq:GPD-momentum-decomp} to compute the quark matrix element $F^{[\Gamma]}_{q/\phi}$ for the composite target.  To do so, it is necessary to go a step further, in the spirit of result found in \cite{Kovchegov:2015zha}, and transform the \textit{spin} decomposition to introduce the spin three-vector $\vec{S}$.  We will do so for the explicit case of spin-$\thalf$ constituents.  This will then allow for a clear identification of the effects that \textit{polarized} intermediate states have in generating composite structure through \eqref{eq:GPD-momentum-decomp}. 

\subsection{Spin Decomposition: Spin-$1/2$ Constituents}

Now let us suppose that the constituents of the composite hadron have spin $1/2$ (or, equivalently, two helicity states).  To fully take advantage of the 3D symmetries in the rest frame of the composite hadron, we need to make a similar decomposition of its spin dependence in terms of the vectors and tensors of 3D rotations.  To bring out the 3D structure of the spin dependence, we change to a Pauli matrix decomposition for the Wigner distribution and the quark GPDs:
%
%
\begin{subequations}
        \label{eq:pauli-decomp}
\begin{align}
    W_{s' s} \big( P_N, b ; P, \Delta \big)
	&=
	W_{\nu} \big( P_N, b ; P, \Delta \big) \:
	[\sigma^{\nu}]_{s' s}
	\notag
	\\
	&\equiv
	W_{unp} \, [\mathbb{1}]_{s' s}
	- \vec{W}_{pol}^{i} \, [\vec{\sigma}^{i}]_{s' s}
	\: , 
    \\ \notag \\
    F_{q/N}^{[{\Gamma}]} ( x' ; P_{N} , \Delta \, ; s' , s )
    &=
    F_{q,\mu}^{[{\Gamma}]} ( x' ; P_{N} , \Delta \, ; s' , s ) \:
    [ \sigma^{\mu} ]_{s s'}
    \notag
    \\
    &\equiv
    F_{q,unp}^{[\Gamma]} 
    \:[ \mathbb{1} ]_{s s'}
    - \vec{F}_{q,pol}^{[\Gamma],j} 
    \:[ \vec{\sigma}^{j} ]_{s s'}
    \notag
    \\
    &\equiv
    A^{[\Gamma]}( x' ; \, P_N,\Delta) \:[\mathbb{1}]_{s s'}
    - \vec{B}_{\perp}^{[\Gamma],\, j}( x' ; \, P_N,\Delta) \: [\sigma_{\perp}^{j}]_{s s'}
    - C^{[\Gamma]}( x' ; \, P_N,\Delta) \:[\sigma^{3}]_{s s'}
    \: . 
\end{align}
\end{subequations}
The Pauli vector $\vec{\sigma}$ is the operator corresponding to the spin three-vector $\vec{S}$, which multiplies the three-vector coefficients $\vec{W}_{pol}, \vec{F}_{q, pol}$.  The vector direction of these functions indicates the vector direction of the spin.

While it is always possible to perform a Pauli decomposition for the eigenspinors of angular momentum projected along any given axis, the interpretation is most straightforward if the spinor polarizations $\sigma$ denote spin projections along the $z$ axis.  For this reason, we work here with the particular spinors defined by Brodsky and Lepage \cite{Lepage:1980fj}, which use spin projections $S^+$ along the light-front $x^+$ axis that correspond to $z$-projections in the rest frame.

The functions that multiply the Pauli matrices in (\ref{eq:pauli-decomp}) are the polarization coefficients of the expansion. Then it is clear from the integrand in (\ref{eq:GPD-momentum-decomp}) and the expansion (\ref{eq:pauli-decomp}) that with the help of the identity $\trace{(\sigma^{\mu}  \sigma^{\nu})} = 2\delta^{\mu\nu}$ the spin dependence decomposes as
\begin{align}
    \frac{1}{2}\sum\limits_{s , s'} W_{s'}^{s} \big(  \vec{P} , \vec{\Delta} ; \,  \vec{P}_{N}  ,  \vec{b}    \big) 
	\cdot F_{q/N}^{[\Gamma]} ( x' ; P_{N} , \Delta \, ; s' , s )
    =
	W_{unp}F_{q,unp}^{[\Gamma]} + \vec{W}_{pol}\cdot\vec{F}_{q,pol}^{[\Gamma]}
    \: .
\end{align}
The master integral for the scalar GPD (\ref{eq:GPD-momentum-decomp}) is now fully understood in terms of two distinct contributions: a polarized and unpolarized channel, which differently manifest the underlying symmetries of the composite hadron:
\begin{align}
    \label{eq:GPD-momentum-spin-decomp}
	F_{q/\phi}^{[\Gamma]} 
    &=
	(1-\xi^{2})\int 
	\frac{ db^{-} \, dP_{N}^{+}  }{ \sqrt{1-\xi_{N}^{2}} }
	\bigg(\frac{1-\alpha}{\alpha}\bigg)
	\int d^{2}b_{\perp} d^{2}P_{N\perp} \, e^{-i[b^{-}\Delta^{+} - \vec{b}_{\perp}\cdot\vec{\Delta}_{\perp} ]}
	\, 
	\big[ 
	W_{unp}F_{q,unp}^{[\Gamma]} + \vec{W}_{pol}\cdot\vec{F}_{q,pol}^{[\Gamma]}
	\big]
    \: .
\end{align}
The structure of the integral (\ref{eq:GPD-momentum-spin-decomp}) suggests for a simple way to perform the integration: by evaluating the boost-invariant Wigner distribution in the target rest frame and performing angular averaging in three dimensions. However, it is not yet clear whether spin-orbit coupling plays a nontrivial role in composite structure or if the polarized term vanishes. One must look deeper into the structure contained in each of the polarization channels and specifically determine the physics arising from the allowed structures in $\vec{W}_{pol}\cdot\vec{F}_{q,pol}^{[\Gamma]}$.  In principle, Eq.~\eqref{eq:GPD-momentum-spin-decomp} is a re-expression of the IA convolution formula that could be equivalently written in terms of a spectral density.  But the Wigner representation \eqref{eq:GPD-momentum-spin-decomp} provides a powerful and manifest way to impose the constraints of \textit{rotational} and \text{discrete} symmetries.

\subsection{Parametrization of the Wigner distribution}
\label{sub:wigner-parame}

Now, we wish to explicitly impose rotational and discrete symmetries on the Wigner density.  As emphasized previously, the key is the boost invariance of the Wigner distribution \eqref{e:W_invariance} which allows us to evaluate it in the rest frame, where its 3D properties are manifest.  The transverse directions are invariant under boosts, so the nontrivial transformations occur to the $z$-components of the three-vectors $\vec{P}_{N}$, $\vec{\Delta}$ and $\vec{b}$.  To exploit the boost invariance of the Wigner density, we simply need to express the rest-frame variables in terms of boost-invariant quantities.


Before setting up the matching, we emphasize that, for the case of GPDs, the initial and final momenta ($p$ and $p'$) of the target hadron are independently on-shell quantities, but the average momentum $P$ is not.  Instead, the value of $P^{+}$ in the rest frame of the target hadron ($\vec{P}\equiv 0$) is determined through the rest energy $(P^{0})$ and the light-front energy $(P^{-})$ obtained from $p$ and $p'$.  Additionally, since the light-front Fourier transform only includes components $b^-, \vec{b}_\bot$, we may consider $b^+ \equiv  0$ the remaining component of the four-vector $b^\mu$.  Then the starting point is the rest-frame quantities
\begin{align}
    P_{RF}^{+}
    &=
    \sqrt{\frac{\frac{1}{4}\vec{\Delta}_{\perp}^{2}+M^{2}}{2(1-\xi^{2})}}
    \quad \text{and} \quad
    b^{-}
    \equiv
    -\frac{1}{\sqrt{2}} (b_{z})^{R.F.}
    \: ,
\end{align}
where we have defined
\begin{align}
    \Lambda^{2} \equiv \frac{1}{4}\vec{\Delta}_{\perp}^{2}+M^{2}
    \qquad , \qquad 
    \lambda^{2}\equiv \frac{1}{4}\vec{\Delta}_{\perp}^{2}+m^{2}
\end{align}
as the effective masses, which are shifted from the true constituent mass $m$ and composite target mass $M$ due to the off-forwardness $\Delta$. The boost-invariant quantities $\alpha, P^+ b^-, \xi$ can be expressed in either frame:
\begin{subequations}
\begin{align}
    \alpha &= \frac{P_N^+}{P^+} = \frac{(P_N^0 + P_{N,z})^{RF}}{\sqrt{2} P_{RF}^+} \: ,
    \\
    P^+ b^- &= P^+_{RF} b^-_{RF} \: ,
    \\
    \xi &= - \frac{\Delta^+}{2 P^+} = - \frac{(\Delta^0 + \Delta_z)^{RF}}{2 \sqrt{2} P^+_{RF}} \: ,
\end{align}
\end{subequations}
which can be solved to express the \textit{equivalent rest-frame three-vectors} for any scattering kinematics in terms of boost-invariant quantities:
\begin{subequations}
\label{eq:z-boosted-quantities}
\begin{align}
    (P_{N,z})^{R.F.}
    &\equiv
    \sqrt{\frac{\Lambda^{2}}{4(1-\xi^{2})}}
    \bigg[ \alpha  - 
    \left(\frac{1-\xi^{2}}{1 -(\xi/\alpha)^{2}}\right)
    \frac{
    \left( \vec{P}_{N\perp} + \frac{1}{2}\vec{\Delta}_{\perp} \right)^{2} - (1-\xi/\alpha)(\vec{P}_{N\perp}\cdot\vec{\Delta}_{\perp}) + m^{2}  }{
        \alpha\Lambda^{2}
    }
    \bigg]
    \\
    (\Delta_{z})^{R.F.}
    &\equiv
    -2\xi\sqrt{\frac{\Lambda^{2}}{1-\xi^{2}}}
    \\
    (b_{z})^{R.F.}
    &\equiv
    -\sqrt{\frac{1-\xi^{2}}{ \Lambda^{2} } }
    (b^{-}P^{+})
    \: .
\end{align}
\end{subequations}
This dictionary allows us to interpret the Wigner function in any boosted frame in terms of its simpler properties in the rest frame.  Note also that the boosted quantities (\ref{eq:z-boosted-quantities}) differ significantly from those found in the TMD results \cite{Kovchegov:2015zha}. In the TMD results, the rest-frame mass is not transformed as in this work, since the momentum that defines the mass ($p^0$) in the TMD results is on-shell. 

For parameterizing the rotational properties of the Wigner function, we will need the scalar products of the various three-vectors:
\begin{subequations}
\begin{align}
    \vec{b}_{RF}^{2} 
    &=
    \vec{b}_{\perp}^{2} + \frac{1-\xi^{2}}{\Lambda^{2}}(b^{-}P^{+})^{2} 
    \\
    (\vec{P}_{N}^{2})_{RF}
    &=    
    \vec{P}_{N\perp}^{2} + \frac{\Lambda^{2}}{4(1-\xi^{2})}
        \bigg[ \alpha  - 
        \frac{
            \Xi^{2}  
        }{
            \Lambda^{2}
        }
        \bigg]^{2}
        \\
    (\vec{P}_{N}\cdot\vec{b})^{2}_{RF}
    &=
    (\vec{b}_{\perp}\cdot\vec{P}_{N\perp})^{2}
    -
    (\vec{b}_{\perp}\cdot\vec{P}_{N\perp})(b^- P^{+}) \left(\alpha - \frac{\Xi^{2}}{\Lambda^{2}}\right)
    +
    \frac{1}{4}(b^- P^{+})^{2}\left(\alpha - \frac{\Xi^{2}}{\Lambda^{2}}\right)^{2}
    \: ,
\end{align}
\end{subequations}
Additionally, there are three new rotational scalars arising from the $\Delta$ phase-space: 
\begin{subequations}
\begin{align}
    (\vec{P}_{N}\cdot \vec{\Delta})_{RF}
    &=
    \vec{\Delta}_{\perp}\cdot\vec{P}_{N\perp} -
    \frac{\xi}{1-\xi^{2}} 
        \bigg[ \alpha\Lambda^{2}  - 
        \Xi^{2} 
        \bigg]
    \\
    (\vec{\Delta}\cdot\vec{b})^{2}_{RF}
    &=(\vec{\Delta}_{\perp}\cdot\vec{b}_{\perp})^{2}
    +
    4\xi (b^{-}P^{+})(\vec{\Delta}_{\perp}\cdot\vec{b}_{\perp})
    +
    4\xi^{2}(b^{-}P^{+})^{2}
    \\
    \vec{\Delta}^{2}_{RF}
    &=
    \vec{\Delta}_{\perp}^{2} + 4\Lambda^{2}\left(\frac{\xi^{2}}{1-\xi^{2}}\right)
    \: ,
\end{align}
\end{subequations}
where 
\begin{align}
    \Xi^{2}\equiv\frac{1}{\alpha}\left(\frac{1-\xi^{2}}{1 -\xi_N^{2}}\right)\left[\vec{P}_{N\perp}^{2} + \lambda^{2} + \xi_N (\vec{P}_{N\perp}\cdot\vec{\Delta}_{\perp})\right]   \: .
\end{align} 
The parameterization in the boosted frame takes nontrivial dependence in terms of the available invariant quantities. Equation (\ref{eq:z-boosted-quantities}) and the inner product invariants presented above are non-trivial results, which lead to the second major theoretical result of this work. 
%
%


The Wigner functions in \eqref{eq:GPD-momentum-spin-decomp} are parameterized in light-front coordinates in the lab frame, appropriate for a high-energy scattering process like DVCS.  However, modeling in these coordinates is not an obvious task.  The natural steps are to start with a model Wigner distribution $W(\vec{\Delta}, \vec{P}_N, \vec{b})$ defined in the rest frame, impose discrete or continuous symmetries, and then change variables to boost-invariant quantities that can be applied in the lab frame using the dictionary (\ref{eq:z-boosted-quantities}).  



In terms of the 3D vectors, the unpolarized part of the Wigner distribution is a rotational scalar that can only depend on scalar products of its arguments:
\begin{align}
	W_{unp} (   \vec{\Delta} , \vec{P}_{N}  ,  \vec{b}   )
	&\equiv
	W_{unp}\big( \vec{b}^{2} , \vec{\Delta}^{2} , \vec{P}_{N}^{2} , (\vec{\Delta}\cdot\vec{P}_{N}) , ( \vec{b}\cdot\vec{\Delta} ) , (\vec{b}\cdot\vec{P}_{N}) , \vec{\Delta}\cdot(\vec{b}\cross\vec{P}_{N}) \big)
	\notag
	\: ,
\end{align}
the arguments can only be constructed from invariant inner products of the available \textit{three}-vectors.
Now, applying discrete symmetries: \textit{parity} and \textit{time reversal} further impose constraints on the structure of the Wigner function and its decomposition. The unpolarized Wigner function must be invariant under the discrete transformations separately. Under \textit{parity} transformations, spatial and momentum coordinates flip sign: $\mathcal{P} \: W(p,s) \rightarrow W(-p,s)$, while for \textit{time} reversal, momentum and spin flip sign: $\mathcal{T} \: W(p,s) \rightarrow W(-p,-s)$. Hence, for the unpolarized function, the kinematic dependence is constrained:
\begin{align}
    \label{eq:wigner-unp-function}
	W_{unp} 
	&\equiv
	W_{unp}\big( \vec{b}^{2} , \vec{\Delta}^{2} , \vec{P}_{N}^{2} , (\vec{\Delta}\cdot\vec{P}_{N}) , ( \vec{b}\cdot\vec{\Delta} )^{2} , (\vec{b}\cdot\vec{P}_{N})^{2} \big)
	\: . 
\end{align}

The spin (polarization) degree of freedom in the polarized Wigner distribution contribution allows for a \textit{vectorial} decomposition from the available \textit{three}-vectors,
\begin{align}
	\vec{W}_{pol}^{i}(   \vec{\Delta} , \vec{P}_{N}  ,  \vec{b}   )
	&=
	w_{1}\vec{b}^{i} + w_{2}\vec{P}_{N}^{i} + w_{3}\vec{\Delta}^{i}
    + w_{4} (\vec{b}\cross\vec{P}_{N})^{i}
	\notag
	\\
	&\qquad 
	+ w_{5} (\vec{b}\cross\vec{\Delta})^{i} + w_{6} (\vec{P}_{N}\cross\vec{\Delta})^{i}
    + w_{7} (\vec{b}\cross(\vec{P}_{N}\cross\vec{\Delta}))^{i}
	\: . 
	\notag
\end{align}
The functions $w_{j}$ are 3D rotational scalars and therefore are constructed from the same arguments as $W_{unp}$ in \eqref{eq:wigner-unp-function}:
\begin{align}
	w_{j} (   \vec{\Delta} , \vec{P}_{N}  ,  \vec{b}   )
	&\equiv 
	w_{j} \big( \vec{b}^{2} , \vec{\Delta}^{2} , \vec{P}_{N}^{2} , (\vec{\Delta}\cdot\vec{P}_{N}) , ( \vec{b}\cdot\vec{\Delta} ) , (\vec{b}\cdot\vec{P}_{N}) \big)
	\: .
	\notag
\end{align}
The (pseudo)vector $\vec{W}_{pol}^i$ is even under parity and odd under time reversal. Then, imposing \textit{parity} covariance for the vectorial decomposition of $\vec{W}_{pol}^{i}$ sets $w_{1}=w_{2}=w_{3}=0$ and $w_{7}=0$. Lastly, \textit{time reversal} forces the polarization index to generate a sign flip that will cancel the sign flip from $\vec{S}$, therefore $w_{6}=0$. In turn, the arguments $(\vec{b}\cdot\vec{\Delta})$ and $(\vec{b}\cdot\vec{P}_{N})$ can only appear in \textit{even} powers, resulting in
\begin{align}
\begin{aligned}
    \label{eq:wigner-pol-decomp}
    \vec{W}_{pol}^{i}
	&\equiv
	(\vec{b}\cross\vec{P}_{N})^{i} w_4 +  (\vec{b}\cross\vec{\Delta})^{i} w_5
    \equiv
    \omega_{1} \: \vec{L}_{P_N}^{i} + \omega_{2}  \: (\Delta\vec{L}^{i})
    \\ \, \\ \mathrm{with} \qquad
	\omega_{j}
	&\equiv
	\omega_{j}\big( \vec{b}^{2} , \vec{\Delta}^{2} , \vec{P}_{N}^{2} , (\vec{\Delta}\cdot\vec{P}_{N}) , ( \vec{b}\cdot\vec{\Delta} )^{2} , (\vec{b}\cdot\vec{P}_{N})^{2} \big)
    \: , 
\end{aligned}
\end{align}
where we have renamed the nonzero coefficient functions in $w_4, w_5$ to $\omega_1, \omega_2)$.

The result for the spin decomposition for the Wigner distribution, after imposing rotational and discrete $\mathcal{P}$ and $\mathcal{T}$ symmetries, implies that the polarized contributions can only come from a particular form of spin-orbit coupling $\vec{L}\cdot\vec{S}$.  This general statement has never been shown before for GPDs in the Impulse Approximation; previous work that stopped at the level of Eq.~\eqref{e:MFexpand1} before proceeding to phenomenology did not impose these highly restrictive symmetry constraints.  An $\vec{L}\cdot\vec{S}$ spin-orbit correlation in the Wigner representation of the type described by $\omega_1$ was previously seen for the case of TMDs, but the correlation described by $\omega_2$ is completely new.  For GPDs,  the angular momentum \textit{transfer} $\Delta\vec{L} = \vec{b}\cross\vec{\Delta}$ to the target can also couple to the constituent spin; this spin-orbit correlation (i) vanishes in the forward limit where $\Delta = 0$ \cite{Kovchegov:2015zha}, (ii) does not appear in previous IA implementations for nuclear GPDs \cite{Guzey:2003,Scopetta:2004,Scopetta:2009,Fucini:2020,Cosyn:2018,Liuti:2005}, and (iii) distinguishes the structure of off-forward distributions from forward parton distributions.  This new coupling channel expands the phase space of allowed structures in composite hadrons and increases the complexity of the convolutions \eqref{eq:GPD-momentum-decomp} or \eqref{eq:GPD-momentum-spin-decomp}.  This finding is the third major theoretical result of this work: 
a new channel of spin-orbit coupling through $\vec{\Delta}$, unique to off-forward processes, which allows for additional composite structure.

\subsection{Unpolarized quark GPD in terms of Wigner distributions in an unpolarized nucleus}
\label{sub:wigner-dist}


The leading-twist GPDs of unpolarized quarks in a spin-$\frac{1}{2}$ hadron are given by the parameterization 
\begin{align}
    \label{eq:spin-half-GPD-spin}
    F_{q/N}^{[{\gamma^{+}}]} (p',s' \, ;  p , s )
    &=
    \frac{1}{2P^{+}}
    \bar{u}(p',s')\bigg[
    H^{q}\gamma^{+}
    + E^{q}\frac{i\sigma^{+\alpha}\Delta_{\alpha}}{2m}
    \bigg]u(p,s)
    \notag \\ \notag \\ &=
    \sqrt{1-\xi_{N}^{2}} \Bigg[
    H_{q}(x',\xi_{N},t) - \left(  \frac{\xi_{N}^{2}}{1-\xi_{N}^{2}} \right)E_{q}(x',\xi_{N},t)  
    \Bigg] \big[\sigma^{0}\big]_{s , s'}
    \notag
    \\
    &\qquad 
    -
    i\frac{\sqrt{1-\xi_{N}^{2}}}{2m}\cdot \epsilon_{\perp}^{i\ell}
    \left( \frac{\vec{\Delta}_{\perp}^{i} + 2\xi_{N}\vec{P}_{N\perp}^{i}}{1-\xi_{N}^{2}}\right)E_{q}(x',\xi_{N},t) \big[\vec{\sigma}_{\perp}^{\ell}\big]_{s , s'}
    \: ,
\end{align}
where we evaluated the Dirac matrix elements using the Brodsky-Lepage spinors (see App.~\ref{ap:polarizations}).
%
Here, $t \equiv \Delta^{2} $ and
Eq.~(\ref{eq:spin-half-GPD-spin}) also provides a clear picture of the polarization channels available for the intermediate spin-$\frac{1}{2}$ states. Importantly, no longitudinally polarized survives to contribute to Eq.~\eqref{eq:spin-half-GPD-spin}; we have separately performed a consistency check verifying this result, see App.~\ref{ap:polarizations}.  



The master formula for the GPDs of an unpolarized quark in an unpolarized composite hadron is obtained by substituting Eqs.~\eqref{eq:wigner-unp-function}, \eqref{eq:wigner-pol-decomp}, and \eqref{eq:spin-half-GPD-spin} into \eqref{eq:GPD-momentum-spin-decomp}: 
\begin{align}
    \label{eq:scalar-gpd-master}
    F_{q/\phi}^{[\gamma^{+}]}(x,\xi,t)
    &=
    (1-\xi^{2}) \int 
	\frac{ db^{-} \, dP_{N}^{+} }{ \sqrt{1-\xi_{N}^{2}} }
	\bigg(\frac{1-\alpha}{\alpha}\bigg)
	\int d^{2}b_{\perp} d^{2}P_{N\perp} \, e^{-i[b^{-}\Delta^{+} - \vec{b}_{\perp}\cdot\vec{\Delta}_{\perp} ]}
    \notag
    \\
	&\times  
	\Bigg\{
    \omega_{0}\sqrt{1-\xi_{N}^{2}}
    \bigg[
    H_{q}(x',\xi_{N},t) - \frac{\xi_{N}^{2}}{1-\xi_{N}^{2}}E_{q}(x',\xi_{N},t) 
    \bigg]
    \notag
    \\
    & 
    +
    \frac{i\omega_{1}}{2m\sqrt{1-\xi_{N}^{2}}}
	\bigg\{
	(b_{z})^{R.F.}\big[ (\vec{\Delta}_{\perp}\cdot\vec{P}_{N\perp})+2\xi_{N}\vec{P}_{N\perp}^{2} \big]
	- \left(P_{N,z}\right)^{R.F.}\big[ (\vec{\Delta}_{\perp}\cdot\vec{b}_{\perp}) +2\xi_{N}(\vec{P}_{N\perp}\cdot\vec{b}_{\perp}) \big]
	\bigg\}E_{q}(x',\xi_{N},t)
    \notag
    \\
    & 
    +
    \frac{i\omega_{2}}{2m\sqrt{1-\xi_{N}^{2}}}
	\bigg\{
	(b_{z})^{R.F.}\big[ \vec{\Delta}_{\perp}^{2} + 2\xi_{N}(\vec{P}_{N\perp}\cdot \vec{\Delta}_{\perp}) \big]
	- \left( \Delta_{z} \right)^{R.F.}\big[ (\vec{\Delta}_{\perp}\cdot\vec{b}_{\perp}) + 2\xi_{N}(\vec{P}_{N\perp}\cdot\vec{b}_{\perp}) \big]
	\bigg\}
	E_{q}(x',\xi_{N},t)
    \Bigg\}
    \: .
\end{align}
We emphasize that the master formula \eqref{eq:scalar-gpd-master}, which makes explicit the mixing channels allowed by symmetry, has never been derived before and is a significant extension beyond calculations that proceed directly from Eq.~\eqref{e:MFexpand1} to phenomenology.  While the $\vec{L} \cdot \vec{S}$ mixing channel described by $\omega_1$ was previously observed in TMDs, the mixing of GPD $E$ with the $\Delta\vec{L} \cdot \vec{S}$ correlation $\omega_2$ is completely new.

The main result in Eq.~(\ref{eq:scalar-gpd-master}) is that the partonic structure of the constituents, described by the GPDs $H_q$ and $E_q$, mixes with nontrivial aspects of the Wigner distribution the constituents within the composite hadron. The angular dependence for the unpolarized channel (arising from the angle between $\vec{b}_{\perp}$ and $\vec{\Delta}_{\perp}$) arises solely from the Fourier transform of the Wigner density described by $\omega_0$.  If the $\omega_0$ has no angular dependence, then the Fourier transform yields the Bessel function $J_0(z)$. On the other hand, the structure of the polarized channel integrand allows for nontrivial angular dependence, even in the case of an isotropic Wigner distribution. In such a case, the angular integral reduces to linear combinations of Bessel functions of different orders. Similarly, the spatial integration of the longitudinal spatial direction $b^-P^+$ results in non-trivial structures for both channels. 

In the unpolarized channel the smearing of $H_q$ and $E_q$ (for nonzero skewness) occurs due to the change of variable of the momentum fraction in the definitions of the composite GPD and due to its convolution with the Wigner function. However, for the polarized channel, only GPD $E_q$ is present and is smeared, but the nontrivial factors multiplying it result in complicated smearing which is also connected to the structure of the Wigner function. Skewness plays a clear role in the overall distribution. Lastly, the integrand allows for a simple/systematic way of identifying which terms drop out after performing angular averaging. In general, it can be shown that the two channels: \textit{unpolarized} and \textit{polarized} give nonzero contributions after integrating
(\ref{eq:scalar-gpd-master}). Similarly, the longitudinally ($\Gamma = \gamma^{+}\gamma_{5}$) and transversely ($\Gamma = i\sigma^{j+}\gamma_{5}$) polarized quark GPD in a scalar target are decomposed into constituent-structure master integrals shown in App.~\ref{ap:polarizations}. The main results in App.~\ref{ap:polarizations} are that the longitudinal polarization distribution goes to zero (as expected), while the transverse distribution has a rich, nontrivial structure. For the scope of this work, we only focus on the unpolarized quark GPD (\ref{eq:scalar-gpd-master}) to demonstrate model calculations. 

It is now instructive to compute the integral in Eq.~(\ref{eq:scalar-gpd-master}) to illustrate the steps that must be taken to fix the kinematics arising from the choice of the Wigner distributions and to show the versatility to easily change models. In practice, a great deal of physics can be learned even for models that are considered ``simple", or serve as a testing ground for different predictions proposed in experiments. Overall, Eq.~(\ref{eq:scalar-gpd-master}) and the comparable results listed in App.~\ref{ap:polarizations} are the fourth major theoretical result found in this work.

\section{Phenomenology: Unpolarized $q$ GPD from Unpolarized Target}
\label{sec:pheno}

Model calculations can be readily performed by choosing a specified model for the Wigner distribution and the constituent nucleon GPDs $H_q$ and $E_q$.  The Wigner distributions can either be computed directly from the first-principles light-front wave functions in some microscopic theory, or they can be modeled by exploiting their semi-classical correspondence.  Simple semi-classical models for the Wigner function include uniform static spheres, Gaussian distributions, and variations on the rigid rotator. The modeling of the distributions $H_q$ and $E_q$ will depend solely on the theory of interest. The list of options is extensive, ranging from the simple quark-target to the scalar-diquark (SDQ) model \cite{Mei_ner_2007,Bhattacharya_2020}. For demonstration purposes, in this work we restrict ourselves to the simplest assumptions: a spherical static distribution and $H_q$ and $E_q$ from the quark target model. This is sufficient to show how a calculation of the scalar target GPD is performed through (\ref{eq:scalar-gpd-master}) and to illustrate the general features arising from model assumptions.

\subsection{Uniform Static Sphere - Distribution}

For demonstration purposes, it is assumed that the Wigner distribution factorizes in terms of spatial and momentum dependent functions,
\begin{align}
    \omega_0 \big(\vec{b} , \vec{P}_{N} , \vec{\Delta} \big)
    &\equiv
    \rho\big(\vec{b}\big)\zeta\big(\vec{P}_{N} , \vec{\Delta}\big) \: . 
\end{align}
Polarized distributions are defined as $\omega_{i} \equiv \beta_i \omega_{0} $ where $\beta_i$ is a polarization coefficient that fixes the amount of $\vec{L}\cdot \vec{S}$ coupling in (\ref{eq:scalar-gpd-master}). In this model, we ignore possible $\vec{\Delta}$ dependence and we choose a static spherical distribution for the spatial and average momentum dependence, independently. This is determined by constraining the phase-space through a hard cutoff (step function):
\begin{align}
    \omega_0 \big(\vec{P}_{N} , \vec{b}\big) 
    &\stackrel{R.F.}{=}
    A_{0} \: \theta\big(R-\abs{\vec{b}}\big) \: \theta\big(P_{N}^{max} - \abs{\vec{P}_N}\big)
    \notag \\ \notag \\ \therefore \quad 
    \omega_0 \big(\vec{P}_{N\perp} , \vec{b}_\bot  \, ; \, P_{N,z}, b_z \big) 
    &\stackrel{R.F.}{=}
    A_{0} \: 
    \theta\big(R - |\vec{b}_\bot|\big) \: 
    \theta\big(P_{N}^{max} - \abs{\vec{P}_{N \bot}}\big) \:
        \notag \\ &\times
        \theta \left(\sqrt{R^2 - |\vec{b}_\bot|^2} - |b_z| \right) \:
        \theta\left( \sqrt{ P_N^{max \: 2} - |\vec{P}_{N \bot}|^2} - |P_{N , z}| \right) \: .  
    \notag \\ \notag \\ \therefore \quad 
    \omega_0 \big(\vec{P}_{N\perp} , \vec{b}_\bot  \, ; \,  P^+ b^- , \alpha \big) 
    &=
    A_{0} \: 
    \theta\big(R - |\vec{b}_\bot|\big) \: 
    \theta\big(P_{N}^{max} - \abs{\vec{P}_{N \bot}}\big) \:
        \notag \\ &\hspace{0.25cm}\times
        \theta\left( \sqrt{ P_N^{max \: 2} - |\vec{P}_{N \bot}|^2} - 
        \frac{\Lambda (\Delta_\bot)}{2}
        \left| \alpha  - 
            \frac{ \vec{P}_{N\perp}^{2} + \lambda^{2} (\Delta_\bot) }
            { \alpha\Lambda^{2} (\Delta_\bot) }
        \right| \right)
        \notag \\ &\hspace{0.75cm}\times
        \theta \left(\sqrt{R^2 - |\vec{b}_\bot|^2} - 
            \left| \frac{P^{+} b^{-}}{ \Lambda (\Delta_\bot)} \right| \right) \: .  
    \label{e:static_sphere_1}
\end{align}
Starting with a model in the rest frame and then converting to boost-invariant coordinates results in a model which is simultaneously boost invariant and under 3D rotational and discrete symmetries.  In contrast, if one were to proceed directly from the light-front wave function representation \eqref{e:MFexpand1} to phenomenology, it would be almost impossible to choose reasonable wave functions which respect the symmetries of the nucleus.  The Wigner representation makes those constraints explicit and systematic.  Here $A_{0}$ is the normalized coefficient and $P_{N}^{max}$ is the maximum momentum carried by the active constituent nucleon. In going from the first to the last line in Eq.~(\ref{e:static_sphere_1}) a substitution of variables was made; from the Cartesian to the boosted light-front coordinates. The normalization $A_0$ is fixed by Eq.~(\ref{e:WignerNorm}) through the unpolarized coefficient, resulting in
\begin{align}
    A_0
    &=
    \frac{1}{\left(\frac{4}{3}\pi R^{3}\right) \left(\frac{4}{3}\pi P_{N}^{max \, 3}\right)
    }
    \: ,
    \notag
\end{align}
which is the inverse value of the space and momentum volume defined by the step functions. The value of $A_0$ in turn defines part of the overall value of the polarized distribution since $\omega_i = \beta_i \omega_0$. 

There is no similar normalization condition that fixes the amount of polarization that may occur. The only constraint present is a \textit{positivity} condition on the number of up versus down polarized constituents. If the condition is 
\begin{align}
    W_{\uparrow, \downarrow} &= W_{unp} \pm \frac{1}{2} W_{pol} \geq 0  
    \notag
    \: ,
\end{align}
then the absolute value of the polarized Wigner function must be less than or equal to twice the unpolarized contribution. This leads to a constraint on the overall polarization coefficient $\beta\big(\vec{P}_{N}, \vec{b}\big) \equiv W_{pol} \big(\vec{P}_{N}, \vec{b}\big) / W_{unp} \big(\vec{P}_{N}, \vec{b}\big)$, which is the ratio of polarized to unpolarized Wigner densities at any point in phase space. It then follows that $+ 2 \geq \beta\big(\vec{P}_{N}, \vec{b}\big) \geq - 2 $, but it is noted that in general $\beta$ fixes the individual $\beta_1$ and $ \beta_2$ which appear for the two cases of spin-orbit coupling in $\omega_1$ and $\omega_2$, as seen from Eq.~(\ref{eq:wigner-pol-decomp}). 
%
%
For models that consider only one $\beta_i$ at a time, this reduces to $| \beta_1 \big(\vec{P}_{N}, \vec{b}\big) \: \vec{b} \times \vec{P}_{N} | \leq 2 $ and $| \beta_2 \big(\vec{P}_{N}, \vec{b}\big) \: \vec{b} \times \vec{\Delta} | \leq 2$,
and since, e.g. $|\vec{b} \times \vec{P}_{N}| \leq |\vec{b}| |\vec{P}_{N}| \leq R P_{N}^{max}$, we can ensure that $\beta_1$ and $\beta_2$ are restricted to the physical range as long as
\begin{subequations}
\label{eq:polarization-bounds}
\begin{align}
    \beta_1 &\leq \frac{2}{R \, P_{N}^{max}}   \: , \\
    \beta_2 &\leq \frac{2}{R \, \sqrt{-t}}  \: .
\end{align}
\end{subequations}
These bounds (\ref{eq:polarization-bounds}) are sufficient to ensure that the positivity constraint is satisfied; they could potentially be relaxed slightly while still satisfying the positivity bound. Also interestingly is the positivity bound on $\Delta\vec{L} \cdot \vec{S}$ coupling $\beta_2$, which is dependent on the momentum transfer $t = \Delta^2$, leading to a weak (no bound) at small or zero $t$, but a restrictive bound on $\beta_2$ at large $t$. All constraints imposed by the Wigner distribution (\ref{e:static_sphere_1}) are fully understood and Eq.~(\ref{eq:scalar-gpd-master}) can be simplified. 

The unpolarized struck quark GPD in a scalar target (\ref{eq:scalar-gpd-master}) decomposes into two polarization channel contributions and greatly simplify in the zero-skewness limit. In the uniform static sphere model (\ref{e:static_sphere_1}), the angular dependence found in the spatial phase-space integrand in (\ref{eq:scalar-gpd-master}) factors from the momentum-space dependence. Resulting in a function that characterizes the smearing of the nucleon GPDs times a different function that dictates the amplitude profile along the $\vec{\Delta}_{\perp}$ axis. The unpolarized channel amplitude is given by
\begin{align}
    \label{eq:Gpd-unp}
    F_{q/\phi}^{[\gamma^{+}],unp}
    &=
    A_{0}[\pi R^{3}][ \pi (P_{N}^{max})^{3}  ]
    \times 
    \Phi_{unp}\big(|\vec{\Delta}_{\perp}|\big)
    \times
    \mathcal{I}_{unp}\big( x , |\vec{\Delta}_{\perp}|  \big)
    \: ,
\end{align}
where the respective functions are defined as
\begin{align}
    \Phi_{unp}\big(|\vec{\Delta}_{\perp}|\big)
    &\equiv
    \frac{2}{R^{3}}\int\limits_{0}^{R} db_{\perp} b_{\perp} J_{0}(b_{\perp}\Delta_{\perp})
    \int\limits_{ - \Lambda \sqrt{R^{2}-|\vec{b}_{\perp}|^{2}}}^{ \Lambda \sqrt{R^{2}-|\vec{b}_{\perp}|^{2}}}  d(b^{-}P^{+})
    =
    \frac{4\Lambda}{|\vec{\Delta}_{\perp}|^{3}R^{3}}
    \big[ \sin(|\vec{\Delta}_{\perp}|R) - |\vec{\Delta}_{\perp}|R \cos(|\vec{\Delta}_{\perp}|R) \big]
    \notag
    \\
    \mathcal{I}_{unp}\big( x , |\vec{\Delta}_{\perp}|  \big)
    &\equiv
    \frac{2}{(P_{N}^{max})^{3}}\int\limits_{0}^{P_{N}^{max}}dP_{N\perp}\, P_{N\perp}
    \int\limits_{\alpha_{min}}^{\alpha_{max}} d\alpha \, \bigg(\frac{1-\alpha}{\alpha}\bigg)
    H_q \left(\frac{x}{\alpha},0,-\vec{\Delta}_{\perp}^{2}\right)
    \notag
    \: .
\end{align}
The limits of integration are determined through the cutoff imposed by the step function from the spherical model. Similarly, for the polarized channel
\begin{align}
    \label{eq:Gpd-pol}
    F_{q/\phi}^{[\gamma^{+}],pol}
    &=
    A_{0}[\pi R^{3}][ \pi (P_{N}^{max})^{3}  ]
    \times 
    \Phi_{pol}\big(|\vec{\Delta}_{\perp}|\big)
    \times
    \mathcal{I}_{pol}\big( x , |\vec{\Delta}_{\perp}|  \big)
    \: ,
\end{align}
and with its corresponding functions defined as:
\begin{align}
    \Phi_{pol}\big(|\vec{\Delta}_{\perp}|\big)
    &\equiv
    -\frac{2}{ R^{3}}\int\limits_{0}^{R}  d b_\bot\, b_\bot^{2} \Delta_{\perp} J_{1}(b_{\perp}\Delta_{\perp})
    \int\limits_{-\Lambda \sqrt{R^{2}-|\vec{b}_{\perp}|^{2}}}^{\Lambda \sqrt{R^{2}-|\vec{b}_{\perp}|^{2}}} 
     d(b^{-}P^{+})
    \notag
    \\
    &=
    \frac{4\Lambda}{|\vec{\Delta}_{\perp}|^{3}R^{3}}
    \bigg[  3R|\vec{\Delta}_{\perp}|\cos(R|\vec{\Delta}_{\perp}|) + \left( R^{2}|\vec{\Delta}_{\perp}|^{2} - 3 \right)\sin(R|\vec{\Delta}_{\perp}|)  \bigg]
    \notag
    \\
    \mathcal{I}_{pol}\big( x , |\vec{\Delta}_{\perp}|  \big)
    &\equiv
    \frac{2}{P_{N}^{max\, 3}}\int\limits_{0}^{P_{N}^{max}} d P_{N\perp} \, P_{N\perp} 
    \int\limits_{\alpha_{min}}^{\alpha_{max}} d\alpha \, 
    \bigg(\frac{1-\alpha}{\alpha}\bigg)
    E_{q}\left(\frac{x}{\alpha},0,-\vec{\Delta}_{\perp}^{2}\right)
    \times 
    \bigg\{
    \beta_{1}
    \frac{\Lambda}{4m\alpha}
        \bigg[ \alpha^{2}  - 
        \frac{
            \vec{P}_{N\perp}^{2} + \lambda^{2}  
        }{
            \Lambda^{2}
        }
        \bigg]
    \bigg\}
    \notag
    \: .
\end{align}
Here $\alpha_{min}$ and $\alpha_{max}$ are functions of $\vec{P}_{N\perp}$ and $\vec{\Delta}_{\perp}$ fixed by solving for $\alpha$ in Eq.~(\ref{e:static_sphere_1}) (see Eqs.~\eqref{eq:alpha-range}). In practice, if $H_q$ and $E_q$ have a simple analytical form, then $\mathcal{I}$ can be calculated in closed form. However, even for simple theories the GPDs are defined through complicated integrals, so numerical calculations are performed. Solving for a model will allow a better understanding of the features arising from the integrals in (\ref{eq:scalar-gpd-master}).

\subsubsection{Helium-4 target ($N \, /\, {}^{4}\text{He}$)}

A natural selection for a composite target to choose as a scalar particle is Helium-4 and its physical parameters fix the boundaries in (\ref{eq:Gpd-unp}) and (\ref{eq:Gpd-pol}). The constituents of Helium-4 are nucleons ($N$), which have spin-$\frac{1}{2}$ and serve as the intermediate constituent-target particle.
The form of the momentum distribution in the model \eqref{e:static_sphere_1}, proportional to $\theta (P_N^{max} - |\vec{P}_N|)$, is characteristic of the zero-temperature filling of states up to a Fermi momentum $k_F = P_N^{max}$.
For simplicity, the internal dimensions of He-4 are modeled after a ``particle in a box", with a maximum length of $R \approx  1.7~\text{fm}$. Then from the uncertainty principle we take $P_{N}^{max} =\hbar c / R \approx  0.12~\text{GeV}$, but a more realistic estimate of the Fermi momentum of ${}^4$He would be larger. Lastly, the mass of a nucleon ($m$) is taken to be equal to $1.0~\text{GeV}$ and the Helium-4 mass $M$ is approximated to be $4.0~\text{GeV}$. The physical parameters then fix the boundary value of the polarization coefficient $\beta_1 \leq 9.8~[\text{GeV}\cdot\text{fm}]^{-1}$, and for the rest of the calculations $\beta_1 \equiv 1~[\text{GeV}\cdot\text{fm}]^{-1}$. The last piece of the calculation is to choose a model for GPDs $H_q$ and $E_q$ that reasonably models the composite structure and that will allow a simple interpretation of how the composite structure influences/modifies the overall distribution (GPD) of the scalar target hadron (Helium-4).
\subsubsection{Quark target model (composite picture): GPD $H_q$ and $E_q$ }

\begin{figure}[t!]
\begin{subfigure}[h]{0.45\linewidth}
\includegraphics[width=\linewidth]{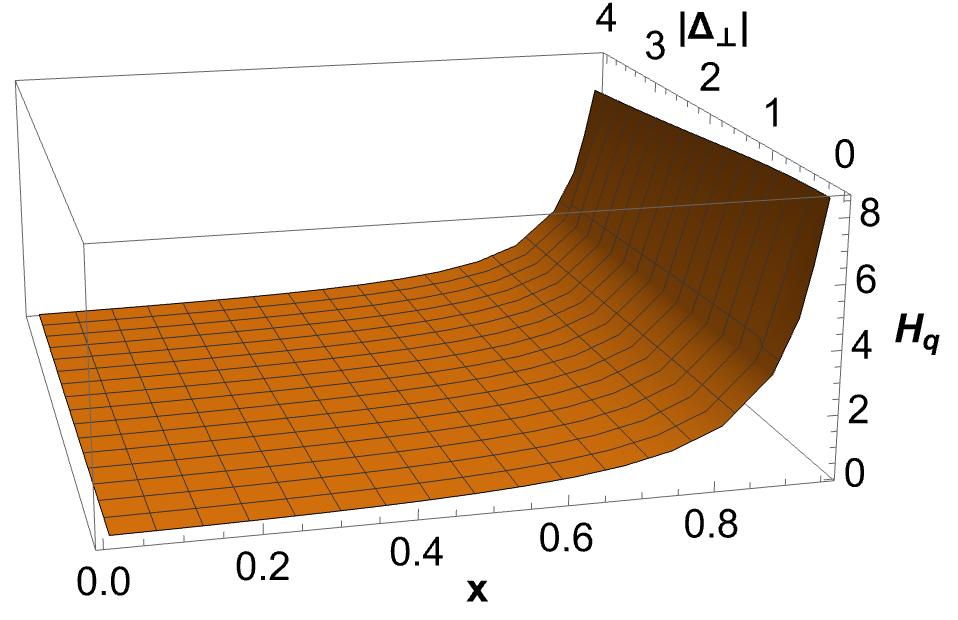}
\end{subfigure}
\hfill
\begin{subfigure}[h]{0.45\linewidth}
\includegraphics[width=\linewidth]{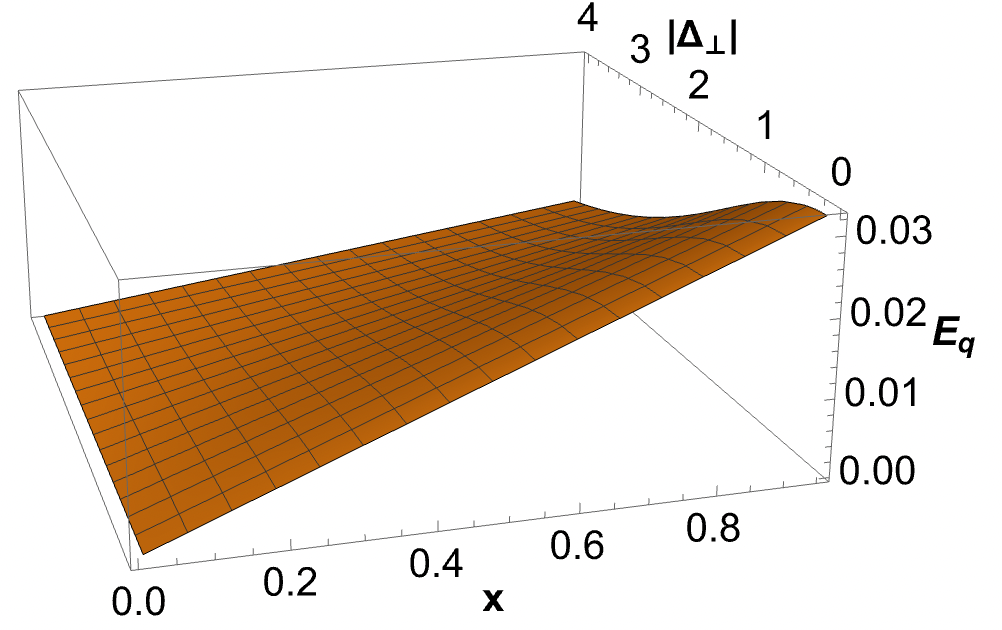}
\end{subfigure}%
\caption{Constituent GPDs $H_q$ (left) and $E_q$ (right) in the quark-target model defined by (\ref{eq:quark-target-model}) for $0<x<1$. }
\label{fig:quark-target-model-Gpds}
\end{figure}

The chosen model is for a quark target, which is obtained from the QCD Lagrangian in perturbative theory. The integrals defining the GPDs $H_q$ and $E_q$ to first order in expansion $\order{g^{2}}$ are taken from \cite{Mei_ner_2007} which hold for $0<x<1$,
\begin{subequations}
\label{eq:quark-target-model}
\begin{align}
    H_q (x, 0, -\vec{\Delta}_\bot^2) 
    &= 
    \frac{4 g^2}{3(2\pi)^3 (1-x)} \int d^2 k_\bot 
    \frac{(1+x^2)[\vec{k}_\bot^2 - \tfrac{1}{4} (1-x)^2 \vec{\Delta}_\bot^2] + (1-x)^4 m^2}{D(x, \vec{\Delta}_\bot, \vec{k}_\bot)}
    \\
    E_q (x, 0, -\vec{\Delta}_\bot^2) 
    &= 
    \frac{8 g^2 x (1-x)^2}{3(2\pi)^3} \int d^2 k_\bot \frac{m^2}{D(x, \vec{\Delta}_\bot, \vec{k}_\bot)}
\end{align}
\end{subequations}
with the (infrared) regularized denominator given by
\begin{align}
    D(x,\vec{\Delta}_\bot,\vec{k}_\bot) &=
    \Big[ \left(\vec{k}_\bot - \tfrac{1}{2} (1-x) \vec{\Delta}_\bot\right)^2 + (1-x)^2 m^2 \Big]
    %
    %
    \Big[ \left(\vec{k}_\bot + \tfrac{1}{2} (1-x) \vec{\Delta}_\bot\right)^2 + (1-x)^2 m^2 \Big] \: .
\end{align}
The use of the quark target model GPDs \ref{eq:quark-target-model} corresponds to modeling the nucleons as single quarks. Again, the idea here to to demonstrate how the modeling works for (\ref{eq:scalar-gpd-master}); for more realistic phenomenology, other models such as the SDQ Lagrangian can be used. In order to compute (\ref{eq:quark-target-model}), the masses are appropriately replaced for those of the target nucleon and $g^2$ is taken to equal unity. The integrals are numerically calculated to illustrate the general properties of the distributions, which are found in Fig.~(\ref{fig:quark-target-model-Gpds}).

Using the quark target model, the GPDs $H_q$ and $E_q$ show different behaviors for the dependence $x$ and $\vec{\Delta}_{\perp}$, as illustrated in Fig.~(\ref{fig:quark-target-model-Gpds}).
There is an increase in amplitude as $x$ approaches one, but it diverges for the case of $H_q$. This can be seen from the overall $x$ dependent factors in (\ref{eq:quark-target-model}) and from the denominator for small asymptotics. This is a reflection coming from the construction of the quark target model, where the initial valence quark at zerth order in perturbation theory introduces a delta function centered at $x = 1$. Hence, the rapid growth of the amplitude as $x\rightarrow 1$ is unique to the QTM. However, qualitatively similar features of a valence quark distribution in a hadron are displayed here.  Those, for instance, would tend to peak around $x \approx 1/3$, rather than $x \rightarrow 1$. The power-law dependence of momentum transfer demonstrated in $H_q \sim \vec{\Delta}_{\perp}^{-2}$ and $E_q \sim \vec{\Delta}_{\perp}^{-4}$ is a reflection of their perturbative (point like) structure. These models extend to large $\vec{\Delta}_{\perp}$ because they have point-like structure at short distances.  It would be expected for this $\vec{\Delta}_{\perp}$ dependence to be greatly shortened (i.e., for a Gaussian or exponential distribution instead of a delta function as an initial condition) in a constituent of finite size. With these features displayed, understanding how the GPDs in (\ref{eq:Gpd-unp}) and (\ref{eq:Gpd-pol}) are transformed is essential in shaping the entire picture of (\ref{eq:scalar-gpd-master}).
%

\begin{figure}[t!]
\begin{subfigure}[h]{0.48\linewidth}
\includegraphics[width=\linewidth]{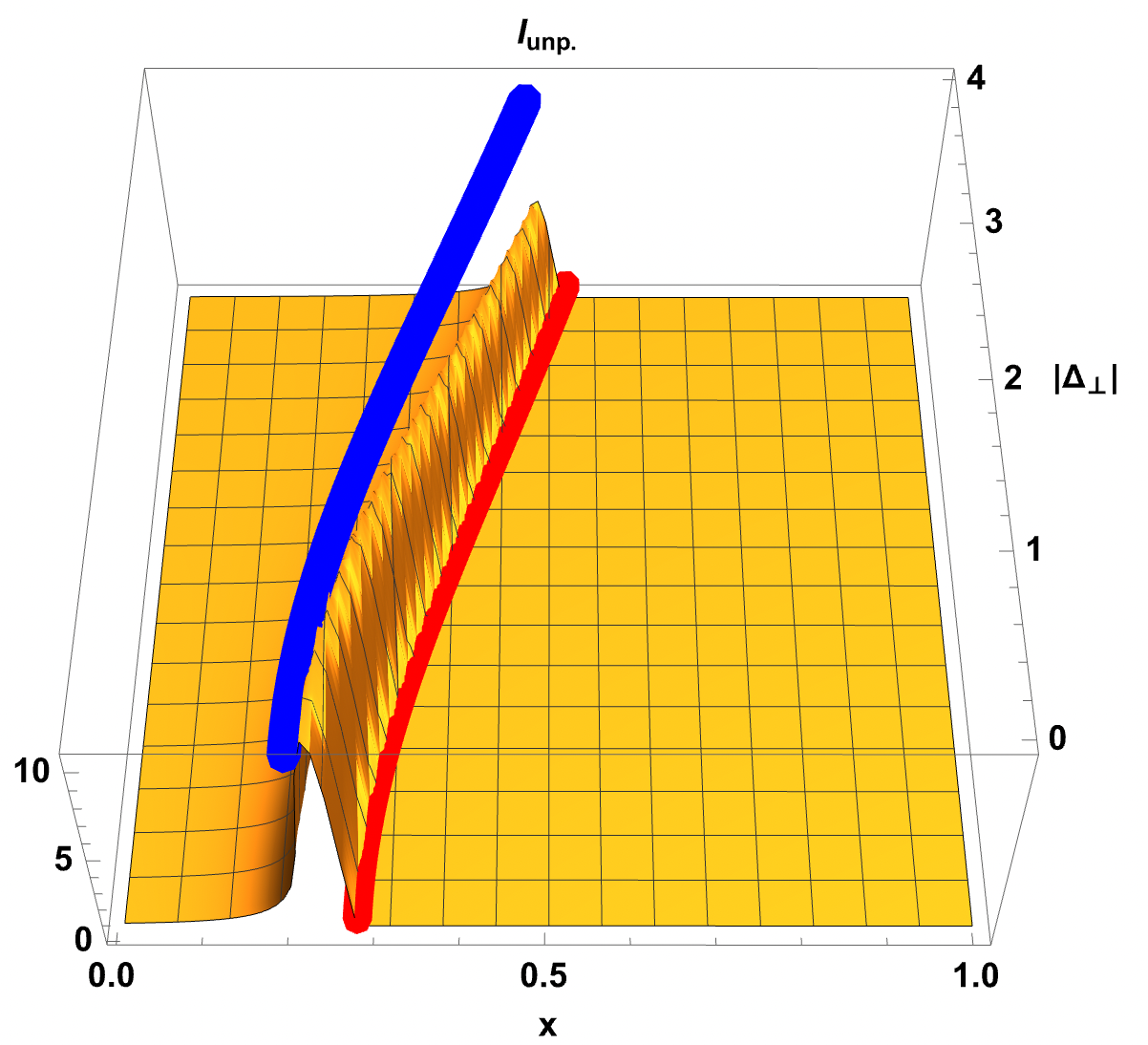}
\end{subfigure}
\hfill
\begin{subfigure}[h]{0.48\linewidth}
\includegraphics[width=\linewidth]{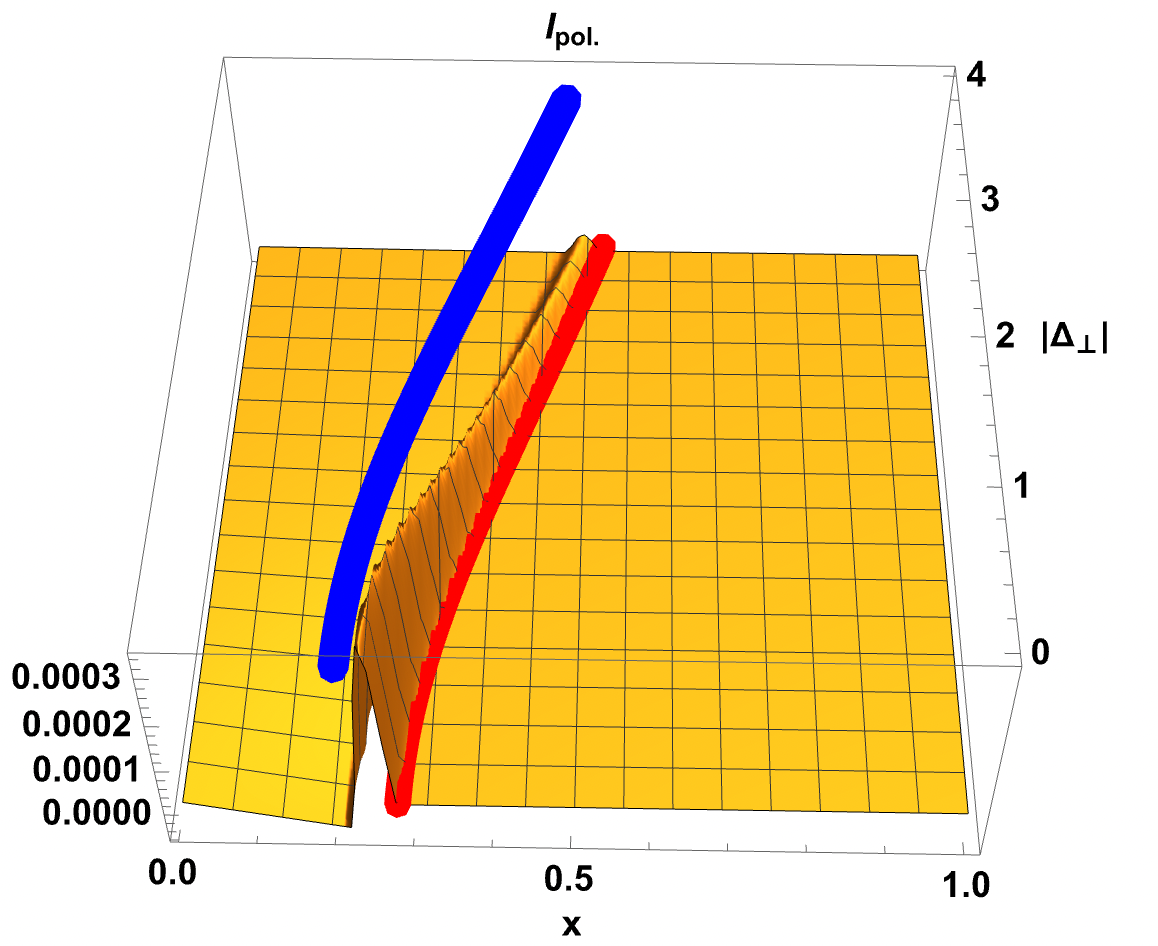}
\end{subfigure}%
\caption{The smearing of GPDs $H_q$ (left) and $E_q$ (right) found in $\mathcal{I}$ corresponding to (\ref{eq:Gpd-unp}) and (\ref{eq:Gpd-pol}), respectively. The \textbf{red} contour represents the maximum value of $x$ for a given $\vec{\Delta}_{\perp}$; above this the distribution is zero. The \textbf{blue} contour represents the critical value of $x$, where the smeared distribution begins to go get cut off.} 
\label{fig:GPD-smearing}
\end{figure}

The integration of $\mathcal{I}_{unp}$ and $\mathcal{I}_{pol}$ in (\ref{eq:Gpd-unp}) and (\ref{eq:Gpd-pol}), respectively, leads to the smearing of $H_q$ and $E_q$ in the longitudinal momentum fraction dependent parameter. One thing to note here is that the integration limits for the $\alpha$ integration changed to $\max{[\alpha_{min},\, x]}$ for the lower limit and $\min{[\alpha_{max},\, 1]}$ for the upper limit. This is a direct result of the longitudinal momentum fraction constraint on the domain arising from the properties of GPDs. The smearing is illustrated in Fig.~(\ref{fig:GPD-smearing}), where $\mathcal{I}_{unp}$ and $\mathcal{I}_{pol}$ are plotted. Many features are illustrated; for one, the smeared GPD $E_q$ falls off faster with $\vec{\Delta}_{\perp}$ arising from the different decomposition momentum fraction structure and its numerator algebra. For large $x$ the distribution goes to zero for some maximum value associated with a fixed $\vec{\Delta}_{\perp}$ value. This is a feature of the phase space boundary $x<1$ of the constituent GPDs being transformed to $x' = x/\alpha < 1$, which imposes a cutoff at a \textit{maximum} value of the momentum fraction $x_{max} = \alpha_{max} (\Delta)$. 
The maximum value of $x$ is illustrated by the red line superimposed on Fig.~(\ref{fig:GPD-smearing}) and the boundary values of $\alpha$ are fixed by Eq.~(\ref{e:static_sphere_1}); which help determine the $x$-dependent features of the distribution. This results in $\alpha_{min} \leq \alpha \leq \alpha_{max}$, which allows for $ x/\alpha_{max} \leq x/\alpha \leq x/\alpha_{min}$, and makes it clear that the fraction $x' = x/\alpha$ will exceed the phase space boundary for at least part of the $\alpha$ integration range when $x$ falls between $\alpha_{min}$ or $\alpha_{max}$.  At $x = \alpha_{min}$ the GPD first begins to become sensitive to the phase space boundary at $x' = 1$; at $x = \alpha_{max}$ all of the allowed phase space has been extinguished.  This leads to the characterization of two distinct features in Fig.~(\ref{fig:GPD-smearing}), a critical $x$-value where the amplitude begins to go to zero ($x_c$) and the maximum value of $x$ where the amplitude equals to zero ($x_{max}$). The $\Delta$ dependent momentum fractions are then given by
\begin{subequations}
\label{eq:alpha-range}
\begin{align}
    x_{max}
    &=
    \alpha_{max}
    =
    \frac{ P_N^{max }  + \sqrt{  P_N^{max \: 2}  + \lambda^{2}  } }{\Lambda}
    \\
    x_{c}
    &=
    \alpha_{min}
    =
    \frac{ - P_N^{max }  + \sqrt{  P_N^{max \: 2}  + \lambda^{2}  } }{\Lambda}
    \\
    \Delta \mathbf{x} 
    &\equiv
    x_{max} - x_{c}
    =
    2\sqrt{\frac{  P_N^{max \: 2}  + \lambda^{2}   }{\Lambda^{2}}}
    \: ,
\end{align}
\end{subequations}
where the quantity $\Delta \mathbf{x}$, denoting the amount of longitudinal momentum smearing, is a signature of composite-structure effects that is directly sensitive to the Fermi momentum $(P_N^{max})$.  In the QTM this manifests in the dramatic peaks seen in Fig.~\ref{fig:GPD-smearing} due to the product of a rapidly increasing valence quark GPD $H_q (x')$, smeared across the kinematic boundary $x' = x/\alpha < 1$ across the finite range of $\alpha$.
Other models may not have the strong peak at $x \rightarrow 1$ that the QTM has, but any notable features of the constituent GPDs will in general be smeared over the longitudinal width $\Delta 
\mathbf{x}$ given in \eqref{eq:alpha-range}.

\begin{figure}[t!]
    \centering
    \includegraphics[width=0.7\linewidth]{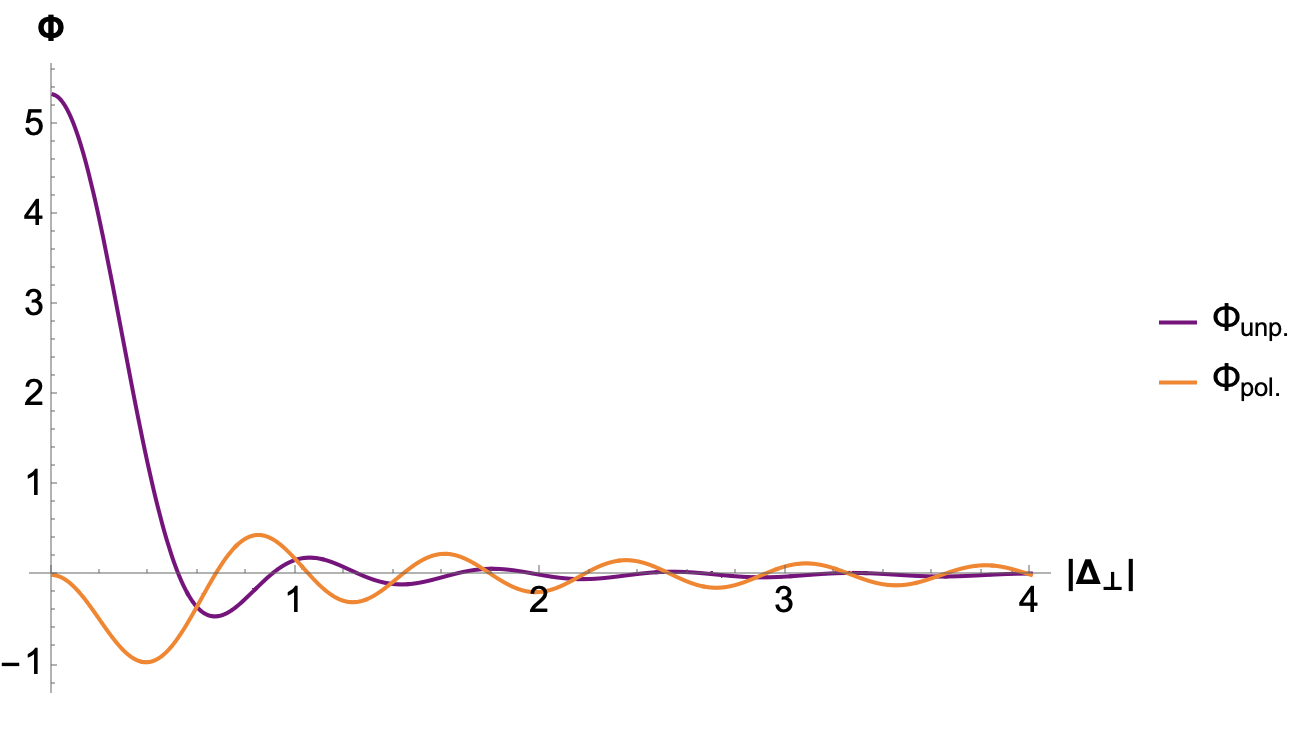}
    \caption{The angular dependence of (\ref{eq:Gpd-unp}) and (\ref{eq:Gpd-pol}) is displayed, which is obtained by integrating over the spatial impact parameter phase-space.}
    \label{fig:angular-depend}
\end{figure}

The other important feature of the unpolarized quark GPD in the composite structure picture (for the model in question) is the factorization of angular dependence ($\Theta$) from the momentum smearing, $\mathcal{I}$. As discussed in Sec.~\ref{sub:wigner-dist} and illustrated by the dependence $\Theta$ in (\ref{eq:Gpd-unp}) and (\ref{eq:Gpd-pol}) for the static sphere model, the angular dependence factors in both channels and the distributions are given by the integral of some factor times a Bessel function. The angular dependence is shown in Fig.~(\ref{fig:angular-depend}), where it is clear that the unpolarized channel results in an enhancement and the polarized channel in a reduction in the overall amplitude, at small $\abs{\vec{\Delta}_{\perp}}$. The angular dependence multiplies the smearing function, which then dictates the modulations along the momentum difference axis. Again, this is a feature that is partially assigned by the uniform static sphere model. 

One of the general characteristics resulting from the momentum cutoff, which fixes the $\alpha$-range associated with the target, is the structure ($\vec{\Delta}_{\perp}$ parameterization) imposed on the $\alpha$ values and the determination of $x_{max}$ and $x_{c}$ as found in Eq.~(\ref{eq:alpha-range}). In general, valence distributions peak at large $x$ and fall-off with increasing $\Delta$, which helps with the overall description of $H_q$ and $E_q$. Other features that are specific include: the amplitude peak of the GPDs in the QTM as $x\rightarrow 1$, the very long range in $\Delta$ and the identification of no enhancement at small $x$. However, it is beyond the scope of this paper to consider a bunch of models at once, but clearly it will be important to explore different choices of model assumptions that will lead to the classification of distinct versus general features. 

\begin{figure}[t!]
\begin{subfigure}[h]{0.45\linewidth}
\includegraphics[width=\linewidth]{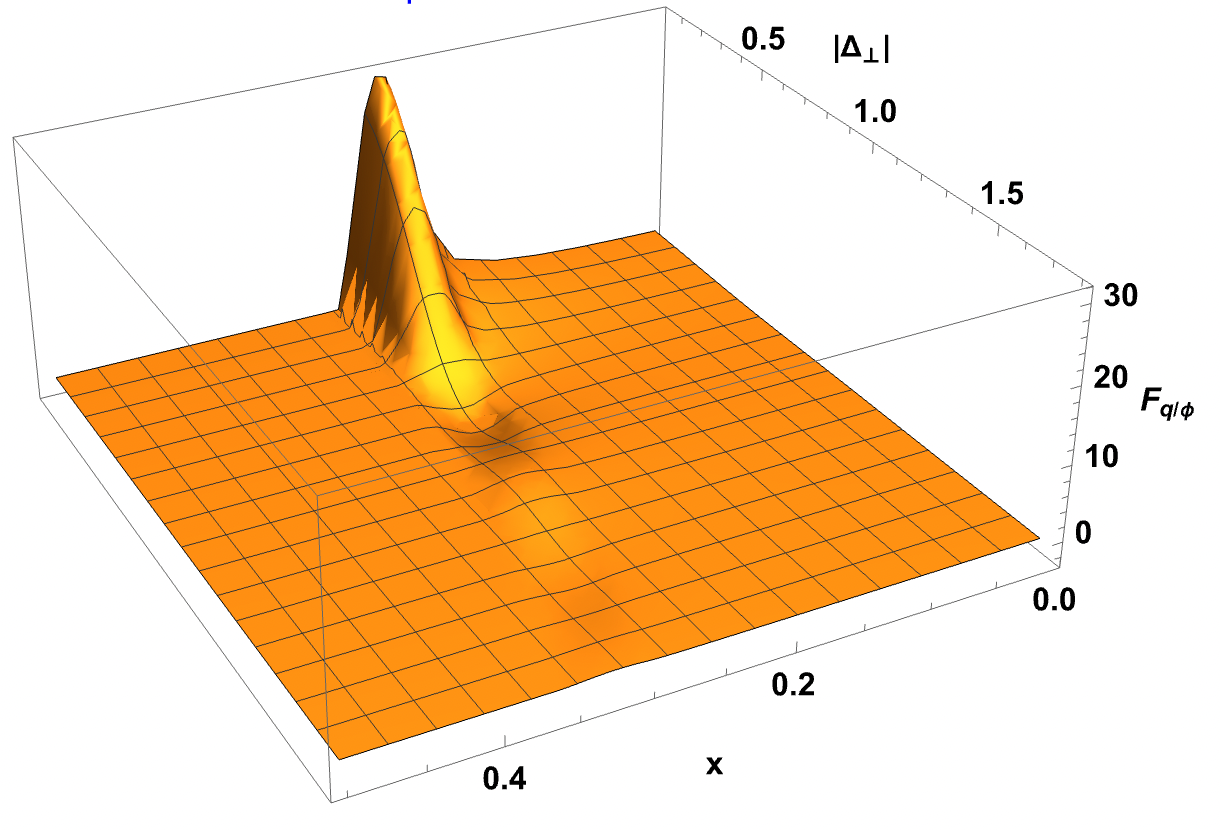}
\end{subfigure}
\hfill
\begin{subfigure}[h]{0.45\linewidth}
\includegraphics[width=\linewidth]{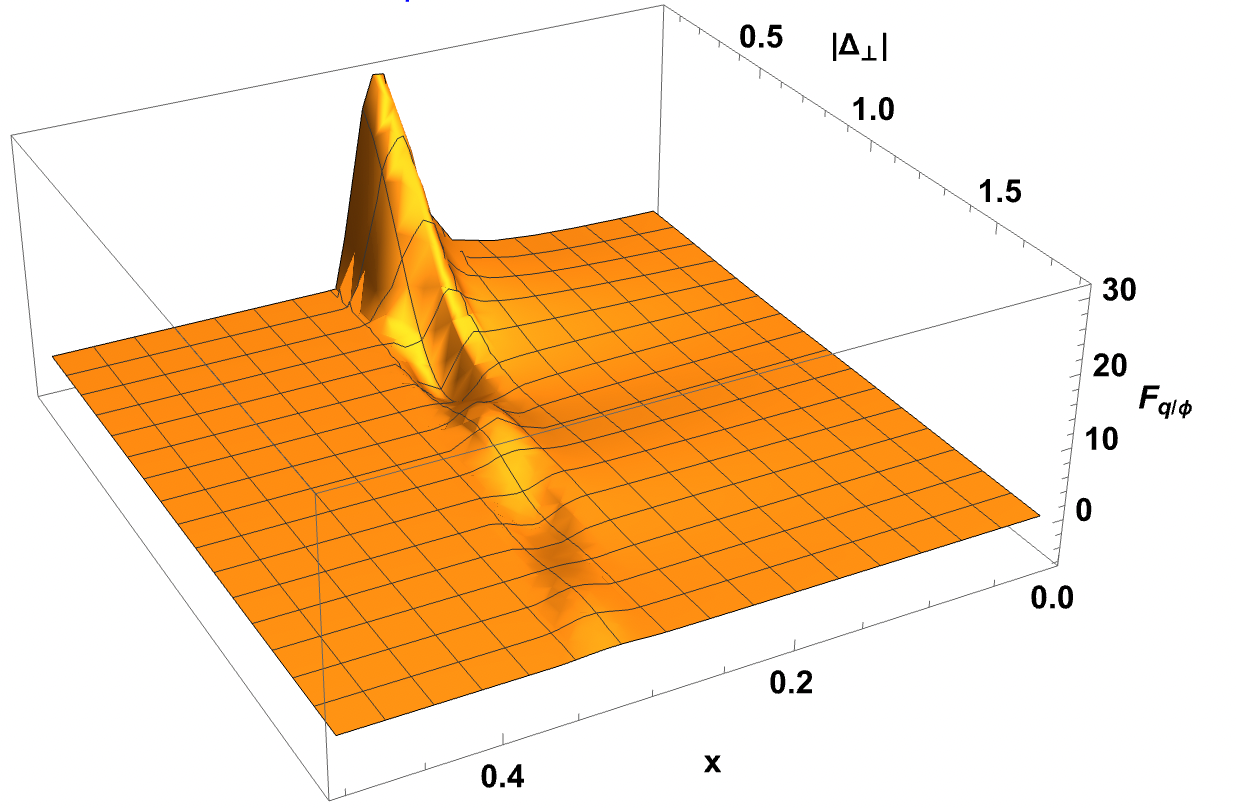}
\end{subfigure}%
\caption{The GPD for an unpolarized struck quark in a scalar ($\phi$), ${}^{4}$He target, is displayed (left) in the composite structure decomposition. The figure on the right displays the effects/role of spin-orbit coupling which is only apparent when $E_q \sim \order{H_q}$.}
\label{fig:GPD-results}
\end{figure}

Now, combining the distinct features allows for an understanding of the overall distribution (\ref{eq:scalar-gpd-master}) for the chosen models. The results are illustrated in Fig.~(\ref{fig:GPD-results}) showing two distinct cases of how the $\vec{L}\cdot\vec{S}$ coupling term affects the overall GPD. The image on the left in Fig.~(\ref{fig:GPD-results}) is the unpolarized quark GPD for Helium-4 with a nucleon composite structure in the uniform static sphere and QTM. In this model $E_q << H_q$ so the influence of the polarized channel (spin-orbit coupling) is not obvious. The general characteristics of the distribution follow from the previous discussions.  We can more clearly see the role/influence of spin-orbit coupling in the overall GPD amplitude if we artificially change the scale of $E_q$ to make it the same order of magnitude as $H_q$: $E_q \sim \order{H_q}$.  The result is illustrated by the image on the right of Fig.~(\ref{fig:GPD-results}), which is the total result (\ref{eq:scalar-gpd-master}) in the QTM with $E_q\sim H_q$ and large $\vec{L}\cdot\vec{S}$ coupling ($\beta_1 \approx 10 $). The profile of the amplitude is a direct result of the angular dependence shown in Fig.~(\ref{fig:angular-depend}).
An interference of the unpolarized and polarized channel is identified in Fig.~(\ref{fig:GPD-results}) determined by the oscillations of the angular profiles displayed in Fig.~(\ref{fig:angular-depend}). These results are a clear indication/identification of the role composite structure plays through spin-orbit coupling in the overall/parent scalar GPD. Part of the overall structure displayed in Fig.~(\ref{fig:GPD-results}) are general and other distinct resulting from the model in question. The key results found are that the structure in the TMD case is also present here but expanded due to the extra momentum transfer variable found in the GPD, not present in the TMD case \cite{Kovchegov:2015zha}. 

\section{Summary and Conclusions}
\label{sec:Concl}

The study of composite structure in GPDs encompasses rich structures that are fundamental to the description of hadrons. In this work, we have generalized the impulse approximation (IA) for GPDs of composite hadrons to a Wigner representation and explicitly enforced the unbroken rotational and discrete symmetries of the nucleus.  The result is a set of master integrals in terms of the Wigner distribution which go far beyond previous results at the level of the wave functions and severely constrain the mixing channels that contribute to composite structure.  An immediate and novel consequence is the nonvanishing contribution of GPD $E$ from polarized intermediate states to the GPDs of an unpolarized composite hadron.  We show that the polarized contribution manifests itself in the form of spin-orbit coupling, as in the case of TMDs \cite{Kovchegov:2015zha}, but we further identify a novel coupling $(\Delta\vec{L}\cdot\vec{S})$ arising from the coupling of spin to the angular momentum \textit{transfer}, which is not present in the case of TMDs.  Lastly, we derived the dictionary between the boost-invariant kinematics appropriate for GPDs and the rest-frame kinematics which exhibit manifest rotational symmetry, finding nontrivial complications due to the off-forward GPD kinematics.  The rich physics of composite structure effects in GPDs is different from but complementary to the physics of composite structure in TMDs, which suggests high potential for application to GTMDs in the future.  The physical picture of composite structure is also not limited only to nuclei, but can also serve as a guide to understanding the role of intermediate degrees of freedom in other hadronic systems. 

To illustrate the physical content of our master formulas, we have also presented an illustrative phenomenological study of composite structure effects on the quark GPDs of a ${}^{4}\text{He}$ nucleus.  We apply the master integral (\ref{eq:scalar-gpd-master}) to a static spherical sphere model of the Wigner distribution (\ref{e:static_sphere_1}) and use a quark target model (\ref{eq:quark-target-model}) for the constituent GPDs.  Although this first exploratory effort is simple, it nonetheless reveals general characteristics of composite structure effects including Fermi motion and spin-orbit coupling and is of direct relevance to DVCS on light nuclei which is being studied in experiments such as E12-17-012 at JLab\cite{Charles2015ALERT, ALERTRunGroup2016}.  The versatility of the master integrals and their suitability for semi-classical modeling make them a natural starting point for phenomenology for such experiments. Equally important, the results found in this paper will help expand the model space for GPDs in the AI-assisted analysis pipeline developed by the EXCLAIM Collaboration.


In the future, it will be instructive to analyze the impact of composite structure on various observables and constraints related to the GPDs.  For one example, GPDs reduce to PDFs in the forward limit, where the effect of the master formula \eqref{eq:scalar-gpd-master} is reduced to a single smearing effect in $x$.  For another, $x$ integrals of the GPDs make connection with hadronic form factors, and a third interesting consideration is the interaction between composite structure and the polynomiality constraint \cite{Diehl2003}.  By generalizing the impulse approximation for composite nuclei and demonstrating its potential for phenomenology, this work constitutes an important step forward in the development and application of the theory of composite structure to hadronic physics.

\section*{Acknowledgments}

This material is based on work supported by the U.S. Department of Energy, Office of Science, Office of Nuclear Physics under Award Number DE-SC0024644 (the EXCLAIM Collaboration) and DE-SC0024560 (MS).  The authors would like to thank M. Engelhardt and W. Horowitz for substantive discussions.

\newpage

\appendix



\section{Definitions of the Relativistic Centers}
\label{app:relativistic-centers}


Herein we summarize the discussion of various choices of the ``center'' of a relativistic system as presented by C. Lorc\'{e} \cite{Lorc__2018,Lorc__2021}.  In general, when the system has a polarization, there can be a subtle frame-dependent distortion of this ``center.''  In the case case at hand, we are interested in a scalar composite target (${}^4 He$), so a priori one need not worry that the center of ${}^4 He$ has been distorted by relativistic effects.  But in the future, we plan to apply this formalism even to polarized composite systems, where consideration of such distortions does become important.  Moreover, even in the case of ${}^4 He$, the question of how to interpret the ``position'' of the polarized constituents is distorted by observer-dependent effects arising from their spin.  The purpose of this Appendix is to discuss those shifts, why they do not distort the calculation presented in the text, and how to interpret the ``dictionary'' \eqref{eq:z-boosted-quantities} from infinite-momentum frame to rest frame.

Consider the \textbf{center of energy} of a system, defined as the average position $x^\mu$, weighted by the timelike component of the energy-momentum tensor $T^{00}(x)$:
\begin{align}
    R_{E}^{\mu}
    &\equiv
    \frac{1}{p^{0}} \int d^{3}x \, x^{\mu} T^{00}(x)
    \, ,
\end{align}
where $p^0 = \int d^3 x \, T^{00}$ is the total energy of the system.  In the rest frame where $\vec{p} = 0$, $R_E$ corresponds to the nonrelativistic center of mass $R_*$:
\begin{align}
    R_{*}^{\mu}
    &\equiv
    R_{E}^{\mu} \:\: \text{if} \:\: \vec{p}=0 \quad (\text{rest frame})
    \, . 
\end{align}
A related but subtly different quantity is the \textbf{center of light-front energy}, defined analogously as the average position $x^\mu$, weighted by the light-front component $T^{++}(x)$ of the energy-momentum tensor:
\begin{align}
    R_{LF}^{\mu}
    &\equiv 
    \frac{1}{p^{+}} \int d^{2}x_{\perp} dx^{-} \, x^{\mu} T^{++}(x)
    \, , 
\end{align}
where $p^+ = \int d^2 x_\bot \, dx^- \: T^{++} (x)$ is the total light-front energy of the system.

As carefully discussed in \cite{Lorc__2018,Lorc__2021}, the centroids $R_E^\mu, R_{LF}^\mu$ do not transform as proper Lorentz four-vectors due to the nontrivial factor of $p^0$ or $p^+$ that appears in the denominator.  As a result, in a general frame, $R_E$, $R_{LF}$, and $R_*$ do not coincide.  In a polarized system, the center of energy $R_E^\mu$ is distorted from the rest-frame center $R_*^\mu$ by a shift which is transverse to both the Pauli-Lubanski pseudovector $\vec{W}$ (i.e., the angular momentum vector) and the system's momentum $\vec{p}$.  As written in Eq.~19 in \cite{Lorc__2021}, the distortion of the center of (instant-form) energy is
\begin{align}
    \vec{R}_{E} - \vec{R}_{*}
    &=
    \frac{ \vec{p}\cross\vec{W} }{ p^{0}m^{2} }
    \, .
\end{align}
In the rest frame, $\vec{R}_E$ coincides with $\vec{R}_*$, and for an unpolarized system ($\vec{W} = 0$) the two always coincide.  But for a polarized system $\vec{W} \neq 0$ and a moving frame $\vec{p} \neq 0$, the center of energy is distorted by a transverse shift relative to the nonrelativistic center $\vec{R}_*$.  Importantly, this finite shift persists even in the ultrarelativistic limit (infinite momentum frame) $\vec{p} = p^0 \, \vec{n}$, $p^0 \rightarrow \infty$.

Likewise, the center of light-front energy $R_{LF}^\mu$ in general is distorted from the position $R_*^\mu$, but in a different way.  As written in Eq.~129 of \cite{Lorc__2018}, the distortion of the center of light-front energy is
\begin{align}   \label{e:LFshift}
    R_{LF}^{\mu} - R_{*}^{\mu}
    &=
    -\frac{ \varepsilon^{\mu\alpha\beta +} W_{\alpha} p_{\beta} }{ m^2 p^{+}  }
    \, .
\end{align}
Again, there is a shift transverse to the polarization $\vec{W}$, but interestingly here the kinematic dependence is reversed.  In the ultrarelativistic limit (infinite momentum frame) $p^+ \rightarrow \infty$, the shift goes to zero; note that $p^+$ cannot enter the numerator due to the Levi-Civita tensor $\varepsilon^{\mu\alpha\beta +}$.  Thus $R_{LF}^\mu$ coincides with $R_*^\mu$ in the infinite momentum frame, whereas in the rest frame $p^+ = p^- = m / \sqrt{2}; \vec{p}_\bot = 0$ the shift remains finite.

\begin{figure}
    \centering
    \includegraphics[width=0.4\linewidth]{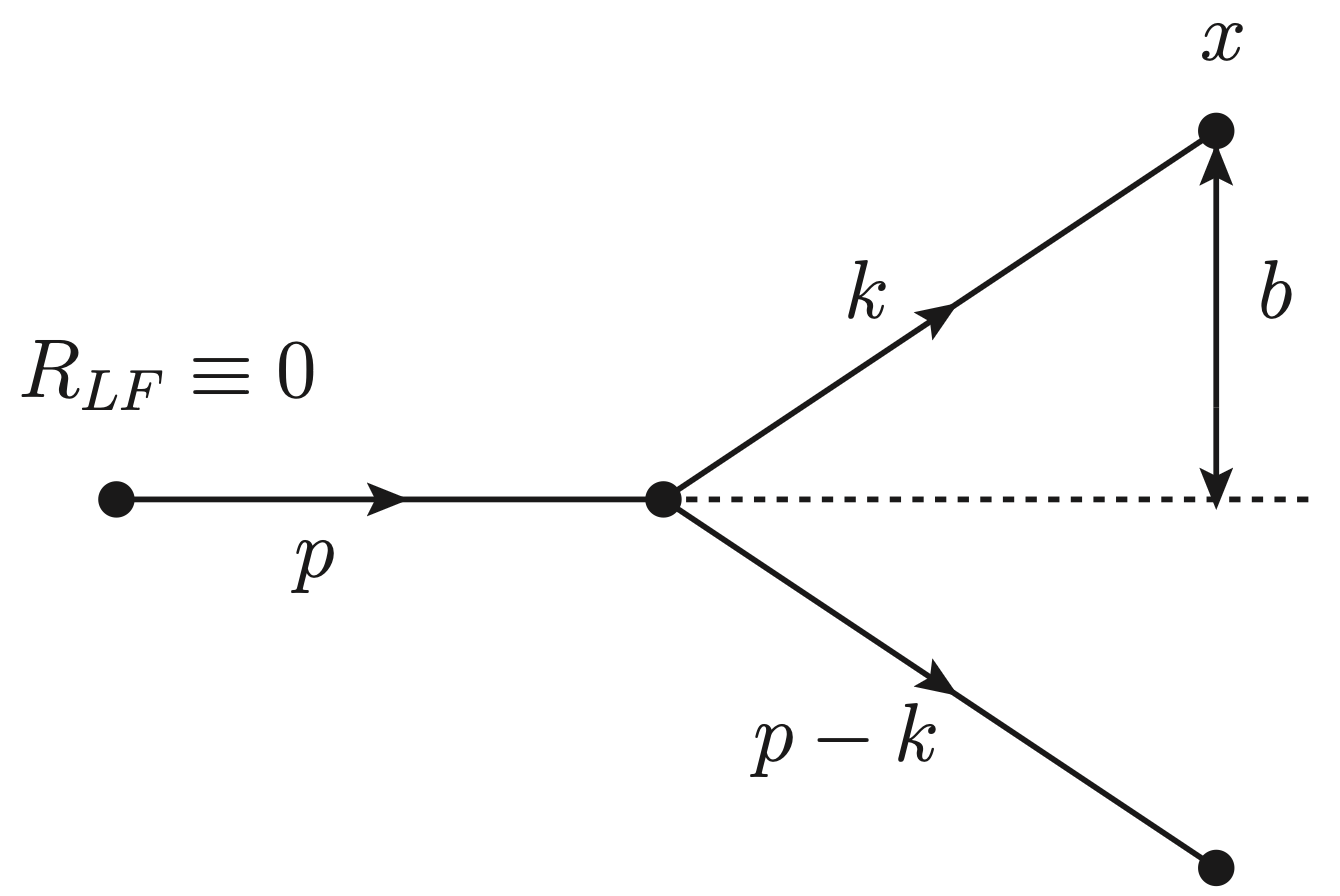}
    \caption{Definition of the impact parameter $b$ as the Fourier transform of a wave function.}
    \label{fig:impact-parameter}
\end{figure}

As carefully discussed in Appendix B of \cite{Lorc__2018}, when one starts with light-front wave functions $\psi(k^+, \vec{k}_\bot)$ and performs a Fourier transform to define a conjugate impact parameter $b^0, \vec{b}_\bot$, 
\begin{align}
    \psi( b^{-} , \vec{b}_{\perp} )
    &=
    \int\frac{dk^{+}}{2\pi}\frac{d^{2}k_{\perp}}{(2\pi)^{2}} \, e^{-ik \cdot b} \tilde\psi(k^{+},  \vec{k}_{\perp} )
    \, . 
\end{align}
the origin of $b^\mu$ corresponds to the center of light-front energy $R_{LF}^\mu$.  Hence, the impact parameter $b^\mu$ as we discuss it in the text corresponds to the displacement of a point $x^\mu$ relative to $R_{LF}^\mu$:
\begin{align}
    b^\mu = x^\mu - R_{LF}^\mu  \: ,
\end{align}
as illustrated in Figure~\ref{fig:impact-parameter}. 

Given that GPDs are defined through \textit{collinear factorization} with asymptotic kinematics $(Q^{2}\rightarrow\infty)$, the matrix element \eqref{eq:quark-gpd-scalar} should be interpreted as being given in the infinite momentum frame.  That is, a GPD is effectively being ``observed'' by a virtual photon moving towards it at the speed of light.  So when we introduce the LFWF and Wigner distributions along with a Fourier transform to impact parameter space \eqref{e:FTdefn}, we may distinguish two different interpretations:
\begin{align}
\begin{aligned}
    b^\mu &\equiv x^\mu - R_{LF}^\mu  \: ,   \\
    \ell^\mu &\equiv x^\mu - R_*^\mu  \: ,   \\
    b^\mu &= \ell^\mu - (R_{LF}^\mu - R_*^\mu)  \: .
\end{aligned}
\end{align}
In the IMF as given, $b^\mu = \ell^\mu$, but we should take caution in interpreting these quantities in the target rest frame.

\begin{figure}
    \centering
    \includegraphics[width=0.7\linewidth]{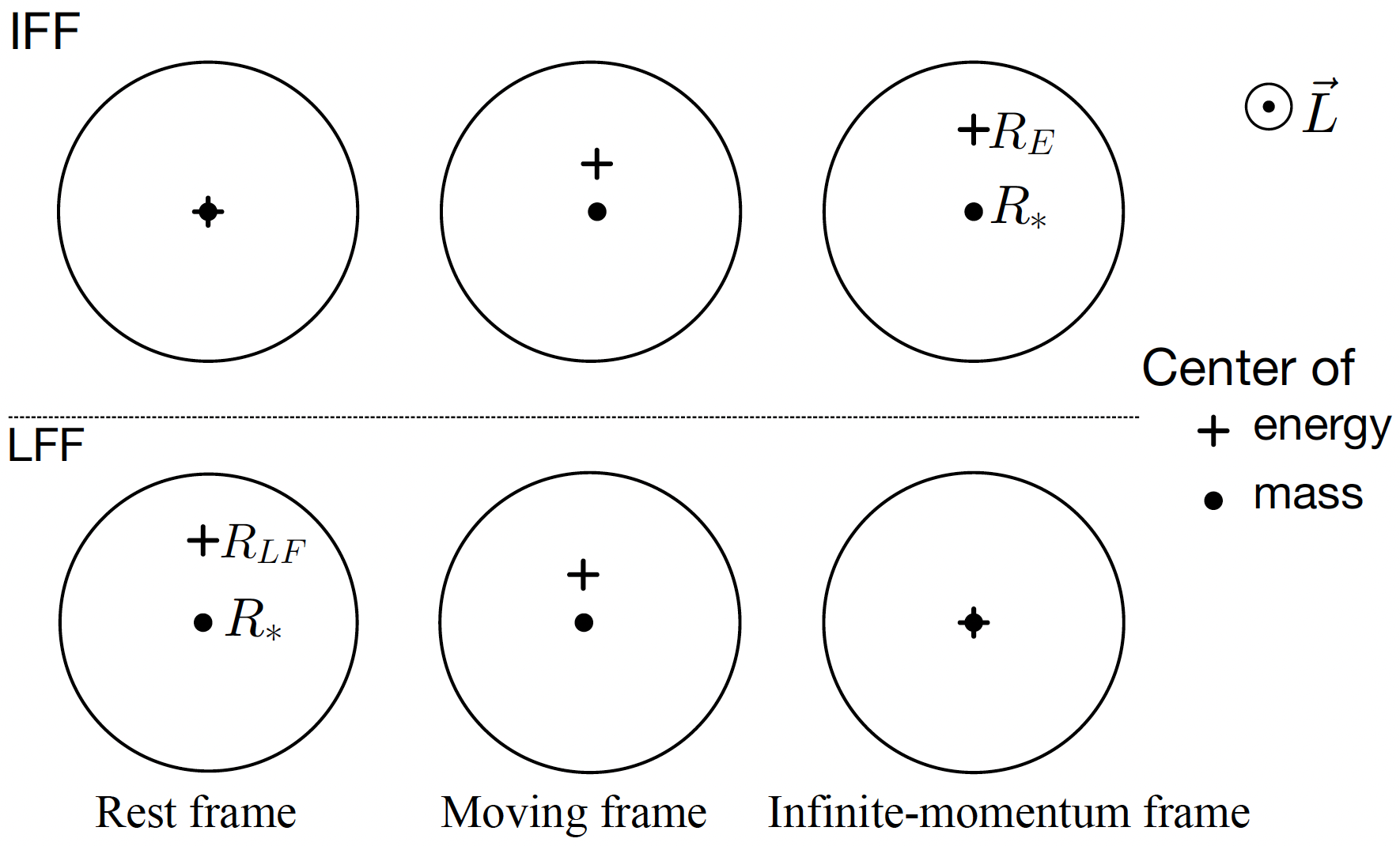}
    \caption{Comparisons of the center of energy and mass for the instant-form formalism (IFF) and light-front formalism (LFF).  Importantly, note that the centers coincide in the rest frame for the IFF but in the infinite-momentum frame for the LFF.}
    \label{fig:interpretation}
\end{figure}
\begin{figure}[t]
    \centering
    \includegraphics[width=0.75\linewidth]{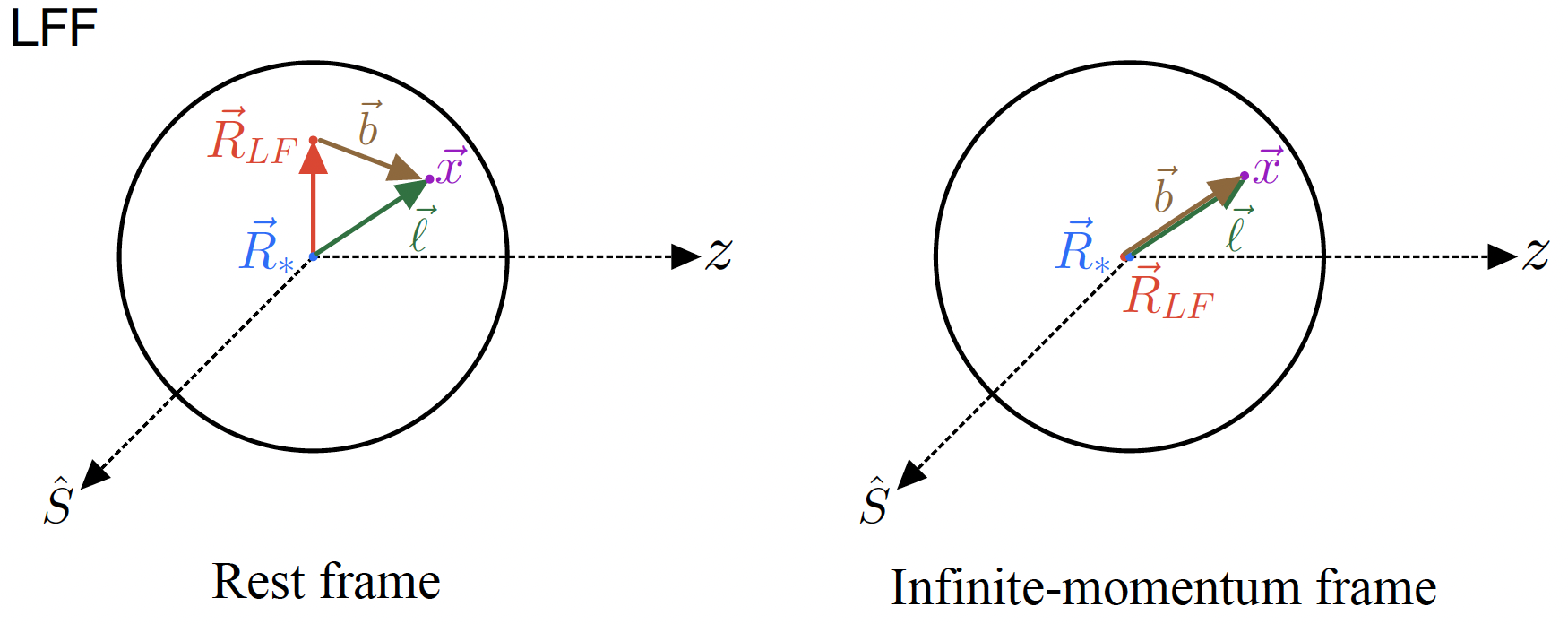}
    \caption{Illustration of the impact parameter $\vec{b}$ with respect to the true four-vector $\vec{\ell}$ defined from the center of mass.}
    \label{fig:3dCenter}
\end{figure}

As drawn in Figs.~\ref{fig:interpretation}-\ref{fig:3dCenter}, in the IMF in which GPDs are defined, the impact parameter $b^\mu$ defined by Fourier transforming the LFWF exactly corresponds to $\ell^\mu$, the displacement relative to the nonrelativistic center $R_*$.  But $b^\mu = x^\mu - R_{LF}^\mu$ is not a proper four-vector, while $\ell^\mu = x^\mu - R_*^\mu$ is.  Therefore, when we form ``boost-invariant products'' like $p^+ b^-$ and interpret them in the rest frame, as in the dictionary \ref{eq:z-boosted-quantities}, what we are really doing is making statements about $p^+ \ell^-$, e.g.
\begin{align}
    P^+ b^- \overset{IMF}{=} P^+ \ell^- = - \sqrt{\frac{\Lambda^2}{1-\xi^2}} \: (\ell_z)^{RF}   \: .
\end{align}
The $\vec{\ell}$ constructed this way gives the three-dimensional displacement vector from the center of mass -- not the displacement vector from the distorted center of light-front energy.

That is, properly evaluating the Fourier transformed impact parameter $b^\mu$ in the rest frame gives $x^\mu - R_{LF}^\mu$, which is the displacement relative to the distorted center.  Due to the distortion, the Wigner distribution would \textit{not} have manifest 3D rotational symmetry about the center of light-front momentum, as a function of the properly boosted $\vec{b}_{RF}$:
\begin{align}
    W \neq W(\vec{b}^2) \: .
\end{align}
But the rest-frame Wigner distribution \textit{does} have manifest 3D rotational symmetry about the center of mass, as a function of the undistorted position vector $\ell$:
\begin{align}
    W = W(\vec{\ell}^2) \: .
\end{align}
So the dictionary \ref{eq:z-boosted-quantities} should not be read as the result of explicitly boosting the non-covariant position $b^\mu$ which includes the distortion of the canonical reference point $R_{LF}^\mu$.  Rather, it translates the 3D properties of $\vec{\ell}$ to the impact parameter $\vec{b}$, which only holds for matrix elements evaluated in the infinite momentum frame.

As a final comment, if one considers the distortion of the canonical center \eqref{e:LFshift} for the \textbf{daughter} particles with $k^\mu \propto p^\mu$, then the shift for both parent and daughter particles vanishes in the IMF, \textit{as long as the longitudinal momentum fraction $x$ is not too small}.  In the simultaneous limit $x \rightarrow 0, p^+ \rightarrow \infty$, it may be possible to have a finite shift in the center of the soft daughter particle, even if the shift in the parent is zero due to the IMF.  That is, at small enough $x$, the parent can be in the IMF while the daughter is not.  This delicate limit $x \rightarrow 0, p^+ \rightarrow \infty$ is outside the usual collinear factorization approach, which assumes a single hard scale $Q \sim p^+ \rightarrow \infty$, while $x$ is assumed to be $\mathcal{O}(1)$.  This interesting limit can be considered further in future work.


\section{Polarized Quark GPDs in Unpolarized Target}
\label{ap:polarizations}

\subsection{Polarized quark GPDs in a spin-$\frac{1}{2}$ target}

The longitudinally polarized quark GPD in spin-$\frac{1}{2}$ target is defined as
\begin{align}
    F_{q/q}^{[\gamma^{+}\gamma_{5} ]}(p,\lambda \, ; p',\lambda ')
    &=
    \frac{1}{2P^{+}}
    \bar{u}(p',\lambda ')
    \bigg[
    \tilde{H}_{q}(x,\xi,t)\gamma^{+}\gamma_{5} + \tilde{E}_{q}(x,\xi,t)\frac{\gamma_{5}\Delta^{+}}{2m}
    \bigg] u(p,\lambda)
    \: ,
\end{align}
and for a transversely polarized quark GPD in spin-1/2 target
\begin{align}
    F_{q/q}^{[ i\sigma^{ j +  }\gamma_{5} ]}(p,\lambda \, ; p',\lambda ')
    &=
    -\frac{i\epsilon_{\perp}^{ij}}{2P^{+}}
    \bar{u}(p',\lambda')
    \bigg[
    H_{q}^{T} i\sigma^{+i} + \tilde{H}_{T}^{q}\frac{P^{+}\Delta^{i} - \Delta^{+}P^{i} }{m^{2}}
    + E_{T}^{q}\frac{\gamma^{+}\Delta^{i} - \Delta^{+}\gamma^{i}}{2m}
    + \tilde{E}_{T}^{q}\frac{\gamma^{+}P^{i} - P^{+}\gamma^{i}}{m} 
    \bigg] u(p,\lambda)
    \: .
\end{align}

\subsection{Spinor products and Pauli decomposition}

Using the Brodsky and Lepage \cite{Lepage:1980fj} spinor convention one gets
\begin{subequations}
\begin{align}
    \bar{u}(p')\gamma^{+}u(p) 
    &=
    2\sqrt{p^{\prime +}p^{+}} [\sigma^{0}]_{\lambda\lambda'}
    \\
    \bar{u}(p')u(p)
    &=
    \sqrt{p^{\prime +}p^{+}} 
    \bigg[ 
    m\bigg( \frac{1}{p^{+}} + \frac{1}{p^{\prime +}} \bigg) [\sigma^{0}]_{\lambda\lambda'}
    - 
    i\epsilon_{\perp}^{ij} \bigg( \frac{\vec{p}_{\perp}^{i}}{p^{+}} - \frac{\vec{p}_{\perp}^{\prime i}}{p^{\prime +}} \bigg) [\sigma_{\perp}^{j}]_{\lambda\lambda'}
    \bigg]
    \\
    \bar{u}(p',\lambda ') \gamma^{+}\gamma_{5} u(p,\lambda)
    &=
    -2\sqrt{p^{\prime +}p^{+}} [\sigma^{3}]_{\lambda\lambda '}
    \\
    \bar{u}(p',\lambda ') \gamma_{5} u(p,\lambda)
    &= 
    \sqrt{p^{\prime +}p^{+}}\bigg[-m\frac{\Delta^{+}}{p^{+}p^{\prime +}}[\sigma^{3}]_{\lambda ' \lambda} + \bigg( \frac{\vec{p}_{\perp}^{i}}{p^{+}} - \frac{\vec{p}_{\perp}^{\prime i}}{p^{\prime +}} \bigg) [\vec{\sigma}_{\perp}^{i}]_{\lambda ' \lambda}\bigg]
    \\
	\bar{u}(p',\lambda')i\sigma^{+i}u(p,\lambda)
	&=
	-2i\sqrt{p^{+}p^{\prime +}} \epsilon_{\perp}^{ij}[\vec{\sigma}_{\perp}^{j}]_{\lambda'\lambda}
	\\
	\bar{u}(p',\lambda')\gamma_{\perp}^{i}u(p,\lambda)
	&= 
	\sqrt{p^{+}p^{\prime +}}
	\bigg\{
	\bigg( \frac{\vec{p}_{\perp}^{i}}{p^{+}} + \frac{\vec{p}_{\perp}^{\prime i}}{p^{\prime +}} \bigg)[\sigma^{0}]_{\lambda'\lambda}
	+
	i\epsilon_{\perp}^{i\ell}
	\bigg[
	\bigg( \frac{m}{p^{+}} - \frac{m}{p^{\prime +}} \bigg)[\vec{\sigma}_{\perp}^{\ell}]_{\lambda'\lambda} + 
	\bigg( \frac{\vec{p}_{\perp}^{\ell}}{p^{+}} + \frac{\vec{p}_{\perp}^{\prime \ell}}{p^{\prime +}} \bigg)[\sigma^{3}]_{\lambda'\lambda}
	\bigg]
	\bigg\}
	\: .
\end{align}
\end{subequations}

\subsection{Master integrals}

In the composite system/picture, the polarized quark GPDs in an unpolarized target are given in terms of the intermediate constituent GPDs, by following similar steps in going from (\ref{eq:GPD-momentum-spin-decomp}) to (\ref{eq:scalar-gpd-master}). Here we present the results for the longitudinally polarized quark distribution in an unpolarized hadron target,
\begin{align}
    F_{q/\phi }^{[\gamma^{+}\gamma_{5} ]}
    &=
    (1-\xi^{2})\int  
	\frac{ db^{-} \, dP_{N}^{+} }{ \sqrt{1-\xi_{N}^{2}} }
	\bigg(\frac{1-\alpha}{\alpha}\bigg)
	\int d^{2}b_{\perp} d^{2}P_{N\perp} \, e^{-i[b^{-}\Delta^{+} - \vec{b}_{\perp}\cdot\vec{\Delta}_{\perp} ]}
    \notag
    \\
    &
    \times 
    \Bigg\{
    \omega_{1}
   \bigg[
   \sqrt{1-\xi_{N}^{2}}\bigg(\tilde{H}_{q}( x' , \xi_{N} , t) + \tilde{E}_{q}( x' , \xi_{N} , t)\frac{\xi_{N}^{2}}{1-\xi_{N}^{2}}\bigg)(\vec{b}_{\perp}\cross\vec{P}_{N\perp})
   \notag
   \\
   &
   -
   (b_{z})^{R.F.}\frac{\xi_{N}}{2m\sqrt{1-\xi_{N}^{2}}} ( \vec{P}_{N\perp}\cross\vec{\Delta}_{\perp}) \tilde{E}_{q}( x' , \xi_{N} , t)
   \notag
   \\
   &\quad 
   - (P_{N,z})^{R.F.} \frac{\xi_{N}}{2m\sqrt{1-\xi_{N}^{2}}} \big[ (\vec{\Delta}_{\perp}\cross\vec{b}_{\perp}) + 2\xi_{N} (\vec{P}_{N\perp}\cross\vec{b}_{\perp})  \big]\tilde{E}_{q}( x' , \xi_{N} , t) 
   \bigg]
   \notag
   \\
   &
    +\omega_{2}
   \bigg[
   \sqrt{1-\xi_{N}^{2}}\bigg(\tilde{H}_{q}( x' , \xi_{N} , t) + \tilde{E}_{q}( x' , \xi_{N} , t)\frac{\xi_{N}^{2}}{1-\xi_{N}^{2}}\bigg)(\vec{b}_{\perp}\cross\vec{\Delta}_{\perp})
   \notag
   \\
   &\quad
   -
   (b_{z})^{R.F.}\frac{\xi_{N}}{2m\sqrt{ 1-\xi_{N}^{2} }}2\xi_{N} ( \vec{\Delta}_{\perp}\cross\vec{P}_{N\perp}  ) 
    \tilde{E}_{q}( x' , \xi_{N} , t) 
   \notag
   \\
   & 
   - (\Delta_{z})^{R.F.} \frac{\xi_{N}}{2m\sqrt{1-\xi_{N}^{2}}} \big[ (\vec{\Delta}_{\perp}\cross\vec{b}_{\perp}) + 2\xi_{N} (\vec{P}_{N\perp}\cross\vec{b}_{\perp})  \big]\tilde{E}_{q}( x' , \xi_{N} , t)
   \bigg]
   \Bigg\}
   \notag
   \\
   &=
   0 \: .
    \notag
\end{align}
The result is equal to zero since every term is proportional to a 2D cross product involving $\vec{P}_{N\perp}$ or $\vec{b}_{\perp}$ linearly, then the angular averaging over these directions cancel due to the parameterization of $\omega_j$. 

Lastly, for a transversely polarized quark in an unpolarized target, $\Gamma = i\sigma^{j+}\gamma_{5} $, we similarly obtain the master integral in terms of constituent spin-$\frac{1}{2}$ GPDs.  The expression is simplest in the zero-skewness limit,
\begin{align}
    \label{eq:trans-pol-master-int}
    F_{q/\phi }^{[i\sigma^{j+}\gamma_{5} ]}
    &\stackrel{\xi\rightarrow 0}{=}
    \int db^{-} \, dP_{N}^{+} 
	\bigg(\frac{1-\alpha}{\alpha}\bigg)
	\int d^{2}b_{\perp} d^{2}P_{N\perp} e^{+i\vec{\Delta}_{\perp}\cdot\vec{b}_{\perp}}
    \Bigg\{
    \omega_{0}
    \bigg\{
    -i\epsilon_{\perp}^{ij}\frac{1}{2m}
    \bigg[
    2\tilde{H}_{T}^{q} \vec{\Delta}_{\perp}^{i} 
    +
    \bigg( E_{T}^{q} \vec{\Delta}_{\perp}^{i} + 2\tilde{E}_{T}^{q} \vec{P}_{N\perp}^{i} \bigg)
    - 2\tilde{E}_{T}^{q}
	\vec{P}_{N\perp}^{i}  \bigg)
	\bigg]
    \bigg\}
    \notag
    \\
    &\quad 
    +
    \omega_{1}
    \Bigg[ 
    \frac{1}{m}
    \tilde{E}_{T}^{q}
	( \vec{b}_{\perp} \cross \vec{P}_{N\perp} )\vec{P}_{N\perp}^{j}
    -
    \bigg\{
	H^{q}_{T} \epsilon_{\perp}^{j i } \vec{P}_{N\perp}^{i}
	-
	\frac{\tilde{H}_{T}^{q}}{2m^{2}}
    \bigg[
    \vec{\Delta}_{\perp}^{2} \epsilon_{\perp}^{j i } \vec{P}_{N\perp}^{i}  
    - (\vec{\Delta}_{\perp}\cross\vec{P}_{N\perp}) \vec{\Delta}_{\perp}^{j}
    \bigg] 
	\bigg\} (b_{z})^{R.F.}
    \notag
    \\
    &\quad
    +
    \bigg\{
	H^{q}_{T} \epsilon_{\perp}^{j i } \vec{b}_{\perp}^{i}
	-
	\frac{\tilde{H}_{T}^{q}}{2m^{2}}
    \bigg[
    \vec{\Delta}_{\perp}^{2} \epsilon_{\perp}^{j i } \vec{b}_{\perp}^{i}  
    - (\vec{\Delta}_{\perp}\cross\vec{b}_{\perp}) \vec{\Delta}_{\perp}^{j}
    \bigg] 
	\bigg\} (P_{N,z})^{R.F.}
    \Bigg]
    \notag
    \\
    &\qquad 
    +
    \omega_{2}
    \Bigg[ 
    \frac{1}{m}
    \tilde{E}_{T}^{q}
	 ( \vec{b}_{\perp} \cross \vec{\Delta}_{\perp} ) \vec{P}_{N\perp}^{j}
    -
    \bigg(
	H^{q}_{T}
	-
	\frac{\tilde{H}_{T}^{q}}{2m^{2}}
    \vec{\Delta}_{\perp}^{2}   
	\bigg) \epsilon_{\perp}^{j i } \vec{\Delta}_{\perp}^{i} (b_{z})^{R.F.}
    \Bigg]
    \Bigg\}
    \: .
\end{align}
For general kinematics, the full result is
\newpage
\begin{align}
    \label{eq:trans-pol-master-int}
    F_{q/\phi }^{[i\sigma^{j+}\gamma_{5} ]}
    &=
    (1-\xi^{2})\int  
	\frac{ db^{-} \, dP_{N}^{+} }{ \sqrt{1-\xi_{N}^{2}} }
	\bigg(\frac{1-\alpha}{\alpha}\bigg)
	\int d^{2}b_{\perp} d^{2}P_{N\perp} \, e^{-i[b^{-}\Delta^{+} - \vec{b}_{\perp}\cdot\vec{\Delta}_{\perp} ]}
    \notag
    \\
    &\times
    \Bigg\{
    \omega_{0}
    (-i)\epsilon_{\perp}^{ij}\frac{\sqrt{1-\xi_{N}^{2}}}{2m}
    \bigg[
    2\tilde{H}_{T}^{q}\bigg(\frac{\vec{\Delta}_{\perp}^{i} + 2\xi_{N}\vec{P}_{N\perp}^{i} }{1-\xi_{N}^{2}} \bigg)
    +
    \bigg( E_{T}^{q} \vec{\Delta}_{\perp}^{i} + 2\tilde{E}_{T}^{q} \vec{P}_{N\perp}^{i} \bigg)
    \notag
    \\
    &\quad
    +
	\bigg(  \xi_{N} E_{T}^{q}  - \tilde{E}_{T}^{q} \bigg)
	\bigg( \frac{2\vec{P}_{N\perp}^{i} + \xi_{N}\vec{\Delta}_{\perp}^{i}}{1-\xi_{N}^{2}} \bigg)
	\bigg]
    \notag
    \\
    &
    +
    \Bigg\{
    \omega_{1}
    \Bigg[ 
    \frac{\sqrt{1-\xi_{N}^{2}}}{2m}
    \bigg( - \xi_{N}E_{T}^{q}  + \tilde{E}_{T}^{q} \bigg)
	\bigg(  \frac{ 2\vec{P}_{N\perp}^{j}+\xi_{N}\vec{\Delta}_{\perp}^{j}}{1-\xi_{N}^{2}} \bigg) ( \vec{b}_{\perp} \cross \vec{P}_{N\perp} )
    \notag
    \\
    & 
    -
    \frac{\sqrt{1-\xi_{N}^{2}}}{2}
    \bigg\{
	2 H^{q}_{T} \epsilon_{\perp}^{j i } \vec{P}_{N\perp}^{i}
	-
	\frac{\tilde{H}_{T}^{q}}{(1-\xi_{N}^{2})m^{2}}
    \bigg[
    \left(\vec{\Delta}_{\perp}+2\xi_{N}\vec{P}_{N\perp}\right)^{2} \epsilon_{\perp}^{j i } \vec{P}_{N\perp}^{i} 
    \notag
    \\
    &\qquad 
    - (\vec{\Delta}_{\perp}\cross\vec{P}_{N\perp}) (2\xi_{N}\vec{P}_{N\perp}^{ j}+\vec{\Delta}_{\perp}^{j})
    \bigg] 
	+ \bigg( \xi_{N}E_{T}^{q} -  \tilde{E}_{T}^{q} \bigg)
	\bigg( \frac{2\xi_{N}}{1-\xi_{N}^{2}}\bigg)\epsilon_{\perp}^{j i } \vec{P}_{N\perp}^{i}
	\bigg\} (b_{z})^{R.F.}
    \notag
    \\
    &
    +
    \frac{\sqrt{1-\xi_{N}^{2}}}{2}
    \bigg\{
	2 H^{q}_{T} \epsilon_{\perp}^{j i } \vec{b}_{\perp}^{i}
	-
	\frac{\tilde{H}_{T}^{q}}{(1-\xi_{N}^{2})m^{2}}
    \bigg[
    \left(\vec{\Delta}_{\perp}+2\xi_{N}\vec{P}_{N\perp}\right)^{2} \epsilon_{\perp}^{j i } \vec{b}_{\perp}^{i} 
    \notag
    \\
    &\qquad 
    -\big[ (\vec{\Delta}_{\perp}\cross\vec{b}_{\perp}) +2\xi_{N}(\vec{P}_{N\perp}\cross\vec{b}_{\perp})\big](2\xi_{N}\vec{P}_{N\perp}^{ j}+\vec{\Delta}_{\perp}^{j})
    \bigg] 
	+ \bigg( \xi_{N}E_{T}^{q} -  \tilde{E}_{T}^{q} \bigg)
	\bigg( \frac{2\xi_{N}}{1-\xi_{N}^{2}}\bigg)\epsilon_{\perp}^{j i } \vec{b}_{\perp}^{i}
	\bigg\} (P_{N,z})^{R.F.}
    \Bigg]
    \notag
    \\
    &
    +
    \omega_{2}
    \Bigg[ 
    \frac{\sqrt{1-\xi_{N}^{2}}}{2m}
    \bigg( - \xi_{N}E_{T}^{q}  + \tilde{E}_{T}^{q} \bigg)
	\bigg(  \frac{ 2\vec{P}_{N\perp}^{j}+\xi_{N}\vec{\Delta}_{\perp}^{j}}{1-\xi_{N}^{2}} \bigg) ( \vec{b}_{\perp} \cross \vec{\Delta}_{\perp} )
    \notag
    \\
    &
    -
    \frac{\sqrt{1-\xi_{N}^{2}}}{2}
    \bigg\{
	2 H^{q}_{T} \epsilon_{\perp}^{j i } \vec{\Delta}_{\perp}^{i}
	-
	\frac{\tilde{H}_{T}^{q}}{(1-\xi_{N}^{2})m^{2}}
    \bigg[
    \left(\vec{\Delta}_{\perp}+2\xi_{N}\vec{P}_{N\perp}\right)^{2} \epsilon_{\perp}^{j i } \vec{\Delta}_{\perp}^{i} 
    \notag
    \\
    &\qquad 
    - 2\xi_{N}(\vec{P}_{N\perp}\cross\vec{\Delta}_{\perp})(2\xi_{N}\vec{P}_{N\perp}^{ j}+\vec{\Delta}_{\perp}^{j})
    \bigg] 
	+ \bigg( \xi_{N}E_{T}^{q} -  \tilde{E}_{T}^{q} \bigg)
	\bigg( \frac{2\xi_{N}}{1-\xi_{N}^{2}}\bigg)\epsilon_{\perp}^{j i } \vec{\Delta}_{\perp}^{i}
	\bigg\} (b_{z})^{R.F.}
    \notag
    \\
    &
    +
    \frac{\sqrt{1-\xi_{N}^{2}}}{2}
    \bigg\{
	2 H^{q}_{T} \epsilon_{\perp}^{j i } \vec{b}_{\perp}^{i}
	-
	\frac{\tilde{H}_{T}^{q}}{(1-\xi_{N}^{2})m^{2}}
    \bigg[
    \left(\vec{\Delta}_{\perp}+2\xi_{N}\vec{P}_{N\perp}\right)^{2} \epsilon_{\perp}^{j i } \vec{b}_{\perp}^{i} 
    \notag
    \\
    &\quad 
    -\big[ (\vec{\Delta}_{\perp}\cross\vec{b}_{\perp}) +2\xi_{N}(\vec{P}_{N\perp}\cross\vec{b}_{\perp})\big](2\xi_{N}\vec{P}_{N\perp}^{ j}+\vec{\Delta}_{\perp}^{j})
    \bigg]
	+ \bigg( \xi_{N}E_{T}^{q} -  \tilde{E}_{T}^{q} \bigg)
	\bigg( \frac{2\xi_{N}}{1-\xi_{N}^{2}}\bigg)\epsilon_{\perp}^{j i } \vec{b}_{\perp}^{i}
	\bigg\} 
    (\Delta_{z})^{R.F.}
    \Bigg]
	\Bigg\}
    \Bigg\}
	\: .
\end{align}
%
%
The structure found here allows for the different intermediate channels to be enhanced with the structure associated with the internal degrees of freedom of the composite particles. A future exercise is then to apply different models to Eq.~{\ref{eq:trans-pol-master-int}} and to obtain/understand the accessible features.


\bibliographystyle{elsarticle-num}
\bibliography{example}






\end{document}